\documentclass[aps,pre,twocolumn,amsmath,amssymb,nofootinbib,superscriptaddress,showpacs,longbibliography]{revtex4-2}
\pdfoutput=1
\usepackage{mathrsfs}
\usepackage{graphicx}
\usepackage{natbib}
\usepackage{latexsym}
\usepackage{amsmath, amssymb, amsthm}
\usepackage{bbold}
\usepackage{mathtools}
\usepackage{xcolor}
\usepackage{comment}
\usepackage{wasysym}
\usepackage[T1]{fontenc}
\def\be{\begin{equation}}
\def\ee{\end{equation}}

\def\bc{\begin{center}}
\def\ec{\end{center}}
\def\bea{\begin{eqnarray}}
\def\eea{\end{eqnarray}}
\newcommand{\avg}[1]{\langle{#1}\rangle}
\newcommand{\Avg}[1]{\left\langle{#1}\right\rangle}
\DeclarePairedDelimiter\ceil{\lceil}{\rceil}

\usepackage[bottom]{footmisc}

\usepackage{epigraph}
\usepackage{lipsum}

\usepackage{float}

\usepackage[normalem]{ulem}

\usepackage{appendix}

\usepackage{hyperref}

\begin{document}

\title{General theory for extended-range percolation on simple and  multiplex networks}

\author{Lorenzo Cirigliano}
\email{lorenzo.cirigliano@uniroma1.it}

\affiliation{Dipartimento di Fisica Universit\`a ``Sapienza”, P.le
  A. Moro, 2, I-00185 Rome, Italy}

\affiliation{Centro Ricerche Enrico Fermi, Piazza del Viminale, 1,
  I-00184 Rome, Italy}

\author{Claudio Castellano}
\email{claudio.castellano@roma1.infn.it }

\affiliation{Istituto dei Sistemi Complessi (ISC-CNR), Via dei Taurini 19, I-00185 Rome, Italy}

\affiliation{Centro Ricerche Enrico Fermi, Piazza del Viminale, 1, I-00184 Rome, Italy}

\author{Ginestra Bianconi}
\email{ginestra.bianconi@gmail.com}
\affiliation{School of Mathematical Sciences, Queen Mary University of London, London E1 4NS, UK}
\affiliation{The Alan Turing Institute, The British Library, 6 Euston Road, London,  NW1 2DB, UK}


\begin{abstract}
Extended-range percolation is a robust percolation process that has relevance for
quantum communication problems. In extended-range percolation nodes
can be trusted or untrusted. 
 Untrusted facilitator nodes are untrusted nodes that can still allow communication between trusted nodes if they lie on a path of distance at most  $R$ between two trusted nodes.  In extended-range percolation the extended-range giant component (ERGC)  includes trusted
nodes connected by paths of trusted and untrusted facilitator nodes.
Here, based on a message passing algorithm, we develop a general theory of
extended-range percolation, valid for arbitrary values of $R$ as long as
the networks are locally tree-like.  This general framework allows us
to investigate the properties of extended-range percolation on
interdependent multiplex networks. While the extended-range nature
makes multiplex networks more robust, interdependency makes them more
fragile. From the interplay between these two effects a rich phase
diagram emerges including discontinuous phase transitions and reentrant phases. The theoretical predictions are in excellent agreement with extensive Monte-Carlo simulations. The proposed exactly solvable model constitutes a fundamental reference for the study of models defined through properties of extended-range paths.
\end{abstract}


\maketitle

\section{Introduction}
\label{sec:introduction}

Percolation
\cite{dorogovtsev2008critical,li2021percolation,lee2018recent,meng2023percolation,artime2024robustness}
is arguably the most fundamental critical phenomenon defined on
networks, as it reflects the connectivity properties and the robustness
of the network on which it is defined. The existence of a giant
cluster is a prerequisite for any collective critical phenomenon
\cite{dorogovtsev2008critical} defined on networks; thus
percolation properties are key to study processes on networks,
such as epidemic spreading phenomena as well as Ising models.

Requiring connectivity might however be a too strong request, and in
many scenarios it is becoming relevant to study percolation problems
in which this requirement is alleviated. This is particularly
interesting in quantum communication \cite{nokkala2024complex,pirandola2020advances,perseguers2010quantum,perseguers2008entanglement}
where noisy data transmission can cause signal degradation~\cite{coutinho2022robustness}. 
Thus quantum networks  require the use of quantum repeaters to extend the range of communication between trusted nodes~\cite{pirandola2019end,zwerger2018long}.  In order to study connectivity of such quantum networks it is therefore important to investigate percolation problems in which the communication between trusted nodes might be allowed also if trusted nodes are not directly connected to each other.
Moreover, for  secure quantum communication, hybrid classical-quantum networks \cite{chen2021implementation} between trusted nodes are often needed, thus requiring the investigation of multilayer \cite{bianconi2018multilayer} percolation properties of these networks.

Here
we focus on extended-range percolation
(ERP), a robust percolation process where nodes may
belong to the same component even if they are not directly connected and we propose a general theory of this model on simple and multilayer networks. This model has been recently proposed and investigated on random networks in Ref. \cite{cirigliano2023extended}.
Also  the lattice version of this model has   attracted significant attention  recently \cite{Malarz2015,
  malarz2020site, malarz2021percolation, xun2020bond, Xun2021,
  zhao2022site, xun2022site}.

There is an increasing interest in percolation problems that do not require the traditional notion of connectivity.
In addition to ERP, several other percolation models have been recently proposed in which percolation is defined through properties of
the shortest paths of the network or through generalized notions of connected components.  These include concurrence   and $\alpha$-percolation \cite{meng2021concurrence,meng2023percolation}, motivated
by quantum communication, shortest path percolation
\cite{kim2024shortest}, motivated by transportation networks, no-exclaves percolation \cite{min2024no}, motivated by the need to enhance network robustness {, and color-avoiding percolation \cite{krause2016hidden,krause2017color}, to reproduce nonuniform vulnerability of nodes to attack or failure}.
Moreover models assuming a nonlocal
  definition of connectivity are raising interest in the last few
  years, also beyond the theory of percolation  with many applications including notably epidemic spreading
  \cite{castellano2020cumulative, cirigliano2022cumulative}. Thus a general theory of ERP on simple and multilayer networks might inspire further research in this  direction.
  
In ERP nodes can be trusted with probability $p$ and untrusted with
probability $1-p$. Untrusted nodes can be involved in the
communication between trusted nodes if they lie on a path between
trusted nodes of length at most $R$.
 {We call these  nodes {\em facilitator} nodes. The extended-range giant component (ERGC) is  formed by all the trusted nodes connected by paths of trusted and facilitator
nodes and a by all the untrusted nodes at distance less than $R$ from these trusted nodes.}  Recently the exact
solution of extended-range percolation on uncorrelated random graphs
with given degree distribution was found for $R$ up to $6$
\cite{cirigliano2023extended}.   {However this solution relies on involved combinatorial definitions which impede a straightforward extension of the approach to treat larger values of $R$ as well as to address generalized percolation problems on multilayer or higher-order networks.}

Here we provide an general theory of ERP for any arbitrary $R$ that  is based on a message passing (MP)
approach   {where the messages have a simple and transparent combinatorial interpretation allowing a deeper theoretical understanding of the model as well as an easier generalization to multilayer networks. } 
The message passing
approach
~\cite{newman2023message,bianconi2018multilayer,mezard2009information}
defines a fundamental distributed computation that applies to a large
variety of critical phenomena and dynamical processes on networks
including percolation, network control, and optimization problems
\cite{liu2011controllability,karrer2014percolation,menichetti2014network,radicchi2017redundant,zdeborova2007phase,hartmann2006phase}.
For standard percolation the MP predicts the size of the giant
component on arbitrary network topology as long as it is locally
tree-like; extensions beyond this approximation are a topic of active
research~\cite{cantwell2023heterogeneous,kirkley2021belief}.
Here we fully develop a message passing theory of ERP that allows us to
predict the size of the ERGC for any value of $R$ and for any
arbitrary network that is locally tree-like.  In this work we also use
this theory to investigate ERP in random graphs with given degree
distribution. We stress that, although the formalism and approach
developed here is markedly distinct from the one formulated in
Ref.~\cite{cirigliano2023extended}, the equations that we obtain for
arbitrary $R$ are equivalent to the ones obtained by the previous work
for $R$ up to $6$.

The general theory for ERP formulated in this work for single networks
is then used to investigate the percolation properties of multiplex
networks under the extended-range framework. Multiplex networks
\cite{bianconi2018multilayer,boccaletti2014structure,kivela2014multilayer}
are formed by $N$ nodes connected via $M$ distinct networks (layers).
Multiplex networks describe a large class
of complex interacting systems where nodes are related by interactions
of different types. Notable examples include interacting
infrastructure and communication networks or biological networks
inside the cell.  The robustness of multiplex networks has raised
significant interest
\cite{buldyrev2010catastrophic,baxter2012avalanche,radicchi2017redundant,bianconi2018multilayer,cellai2016message,min2015link},
as interdependencies between the layers can lead to an increased
fragility of the system, thus providing a framework to model cascades
of failure events propagating across the layers of the multiplex
network. In particular in Ref.~\cite{buldyrev2010catastrophic} a
multiplex interdependent percolation (MIP) problem has been
defined. The order parameter of this model is given by the size of the
mutually connected giant component (MCGC) formed by nodes (directly)
connected to each other by at least one path on each layer. The MCGC
emerges exhibiting a discontinuous hybrid phase transition, and its critical
properties reflect the increased fragility of the system.

In this work we investigate the trade-off between the effect of
interdependencies {present in multiplex networks} that increase the
fragility of the multiplex networks, and the extended-range mechanism that facilitates communication between otherwise
disconnected nodes. To this end, we formulate the Multiplex Extended-Range Percolation (MERP) whose order parameter is the size of the Multiplex Extended-Range Giant Component (MERGC).
We introduce a notion of interdependency for
trusted nodes. However, since untrusted nodes play a significant role in extended-range percolation, the definition of the model is not complete if we do not specify whether untrusted nodes are interdependent. This leads to two variants of the MERP (version $V=1$ and version $V=2$). In version $V=1$ of MERP an untrusted node belongs to the MERGC if it belongs to the ERGC {\em in each layer}. Thus this version imposes interdependencies for untrusted nodes. In version $V=2$ instead, it is sufficient that an untrusted node is part of the ERGC in {\em at least one} layer to belong to the MERGC,
i.e., we do not impose interdependencies for untrusted nodes.

We provide an exact solution of both versions of MERP on uncorrelated random multiplex networks with arbitrary degree distributions. Our analysis of the critical phenomena of MERP is based upon analytical predictions of the phase diagram, supported by extensive Monte Carlo simulations. We reveal the important interplay between the effect of interdependecies and of the extended-range mechanism.
In particular, the study of version $V=2$ of MERP shows a highly nontrivial phenomenology with a reentrant
discontinuous hybrid phase transition, observed in multiplex networks where the MCGC does not exist.

This paper is structured as follows:
In Sec. II we define extended-range percolation on simple networks and we present a general message-passing theory to predict the size of the ERGC on arbitrary tree-like networks. In Sec. III we formulate a general theory of extended-range  percolation on random networks, valid for any arbitrary choice of the range $R$ and we compare our theory with Monte Carlo simulations. In Sec. IV we formulate the two versions of MERP (version $V=1$ and version $V=2$). We provide a general analytic theory to calculate the MERGC and we discuss the critical properties of MERP, including the presence of a characteristic reentrant phase transition in the version $V=2$ of this model. Our theoretical predictions are in excellent agreement with Monte Carlo simulations. Finally in Sec. V we provide the concluding remarks.
The paper is enriched with Appendices, providing details of the derivation of the message passing theory for ERP, explicit equations for ERP and MERP for finite values of $R$, and the discussion of the equivalence of  the equations of ERP for $R\leq 4$ with the equations derived in Ref. \cite{cirigliano2023extended}.

\begin{figure}[!htb]
   \centering
   \includegraphics[width=0.48\textwidth]{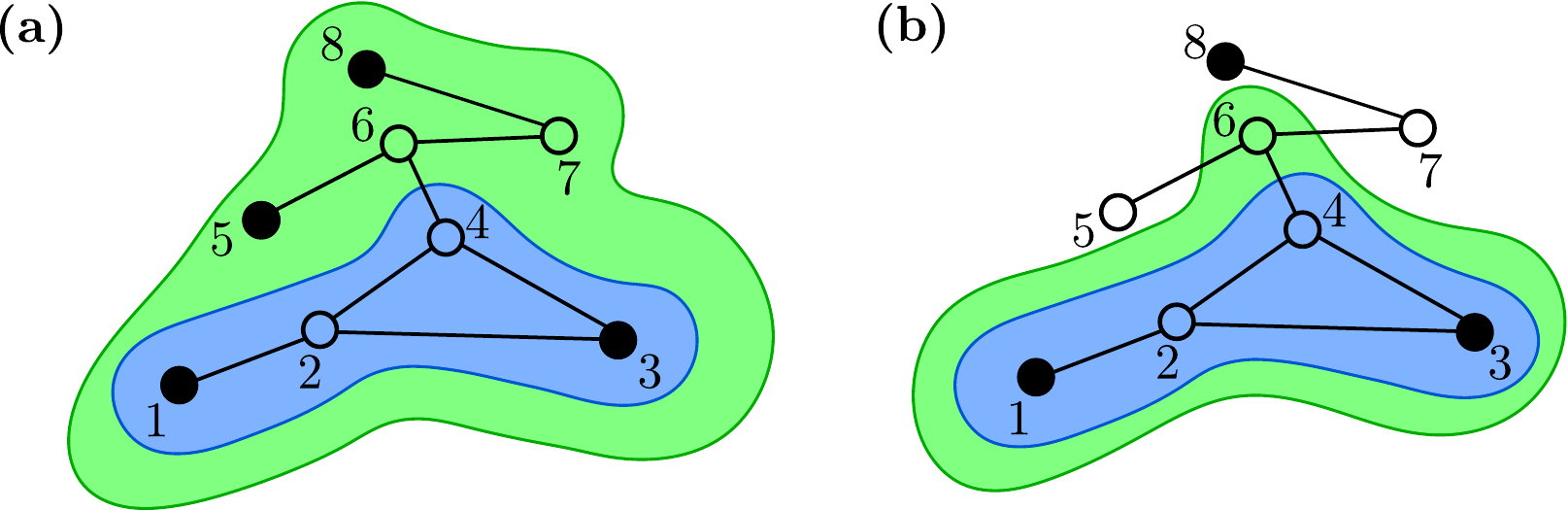}\caption{Schematic representation of extended-range percolation on networks. Trusted and untrusted nodes are represented by filled and empty circles, respectively. The  extended-range giant component (ERGC) of range $R=2$ (smaller, blue component) and $R=3$ (larger, green component) are highlighted. The two networks in (a) and (b) have the same $R=2$ ERGC, but different $R=3$ ERGC. For $R=2$, node $4$ belongs to the ERGC even if it is not a facilitator. For $R=3$, node $6$ is a facilitator in (a) but not in (b); neverthless, in both cases it is a part of the ERGC. Node $7$ instead is in the ERGC in (a) -- since it is a facilitator -- but not in (b).}
   \label{fig:schematic_ERP}
\end{figure}

\section{Extended-range percolation and message passing approach}
Consider a network $\mathcal{G}=(V,E)$ of $N=|V|$ nodes where nodes
are either untrusted  or trusted. Communication between
two trusted nodes takes place only if there is at least one  walk connecting them including trusted and untrusted nodes and having  
at most $R-1$ consecutive untrusted nodes.  Note that
we require the existence of a {\em walk}  and not a {\em path}. For
instance, nodes $3$ and $8$ in Fig.~\ref{fig:schematic_ERP}(a) can communicate for
$R=3$ even if the (shortest) path $\{3 \to 4 \to 6 \to 7 \to 8 \}$
connecting them contains three consecutive untrusted nodes,
thanks to the existence of the walk $\{3,4,6,5,6,7,8\}$ having at most two consecutive untrusted nodes because of the presence of the trusted node $5$.

 {Strictly speaking, communication is among trusted nodes, but it is clear that untrusted nodes play a crucial role in this process. In particular, untrusted nodes lying on paths between two trusted nodes of length at most $R$. We call such untrusted nodes \textit{facilitators}, because they are allowing for the communication of distant trusted nodes. For
instance, node $2$ is a facilitator for $R=2$ and $R=3$ in Fig.~\ref{fig:schematic_ERP}, nodes $4,6,7$ are facilitators for $R=3$ in Fig.~\ref{fig:schematic_ERP}(a) but not in (b).}
Given this way to communicate, we  {define the extended-range giant  connected component (ERGC) as the giant component formed by  trusted nodes connected by paths including exclusively  trusted or untrusted facilitator nodes and by all the untrusted nodes at distance less than $R$ from these trusted nodes.} This is equivalent to say that any pair of trusted nodes in the ERGC is connected by at least a walk formed by trusted and untrusted nodes including at most $R-1$ consecutive untrusted nodes (facilitators).  {Moreover the untrusted nodes in the ERGC include all the facilitator nodes and all the untrusted nodes that are not facilitators but are at distance less than $R$ from the trusted nodes in the ERGC.}  For a schematic representation of the ERGCs see Fig. \ref{fig:schematic_ERP}.

 {The algorithm to evaluate the ERGC is a generalization of the Depth-First Search algorithm.
We assign to each node $i$ of the network a pair of variables  $(q_i,r_i)$. Initially no node is assigned to a cluster, i.e. every node has $q_i=0$. Moreover all trusted nodes have $r_i=0$ and all untrusted nodes have $r_i=R$. At the end of the algorithm $q_i=q\in\{1,2,\ldots,Q\}$  indicates to which of the $Q$ clusters node $i$ belongs and $r_i$  indicates the distance (if smaller than $R$) from the closer trusted node. 
The algorithm for detecting the ERGC is  the recursive Extended-Range Depth First Search ERDFS algorithm.
Starting from a an arbitrary trusted (unvisited) seed node for cluster $q$ (initially taken to be $q=1$), we set $q_i=q$ and $r_i=0$ and we call the ERDFS algorithm.\\
The recursive ERDFS algorithm is defined as follows.
Starting from a  node $n$ with $q_n=q$ and $r_n=r$, ERDFS implements (1) and (2) defined as: 
\begin{itemize}
\item[(1)]
If $r<R$ the algorithm visits all the neighbour trusted nodes $m$ unvisited previously. For each of these nodes the algorithm sets  $r_m=0$ and $q_m=q$, and the ERDFS  is recursively iterated.
 \item[(2)] If $r<R-1$ the algorithm visits all the neighbour untrusted nodes $m$ with $r_m>r+1$.
 For each of these nodes the algorithm sets $r_m=r+1$ and $q_m=q$, and the ERDFS  is recursively iterated.
 \end{itemize}
When there are no more nodes to visit, the ERDFS stops and all the trusted and untrusted nodes with $q_i=q$ form the ERP cluster $q$. If there are unvisited trusted nodes we choose a new unvisited trusted seed node for cluster $q+1$ and iterate the procedure until  and all the trusted nodes are assigned to a cluster and we have $Q$ clusters (where $Q$ is determined by the algorithm).
The ERGC is the giant cluster  $q=q^{\star}$ formed by the trusted nodes with $r_i=0$ and the untrusted nodes with $r_i<R$ all having $q_i=q^{\star}$.}\\

Here we develop a message-passing (MP) theory to predict the size of the ERGC on any locally tree-like network.
The general MP equations describing the problem with a given configuration of trusted and untrusted nodes  are explained in detail in Appendix \ref{ApMP}.
Here we assume that nodes are trusted with probability $p$ and untrusted with probability $1-p$, and the equations we obtain follow from the more general ones, see Appendix \ref{ApMP}.
We can express the size of the ERGC using two sets of cavity messages $\hat{\sigma}_{i\to j}^r$ and $\hat{\omega}_{i\to j}^r$, for $1\leq r\leq R$. The generic message $\hat{\sigma}_{i\to j}^{r}$ sent from node $i$ to node $j$ indicates the probability that node $i$ is in the ERGC and it is at distance $r-1$ from the closest trusted node when the link $(i,j)$ is removed. The generic message $\hat{\omega}_{i\to j}^{r}$ sent from node $i$ to node $j$ indicates the probability that node $i$ is not in the ERGC and it is at distance $r-1$ from the closest trusted node when the link $(i,j)$ is removed.

Our  message passing equations (see Appendix \ref{ApMP} for a detailed derivation) dictate that the messages  $\hat{\sigma}_{i\to j}^r$  obey
\begin{widetext}
\bea
\hat{\sigma}_{i\to j}^1&=&p\left[1-\prod_{\ell\in \partial i\setminus j}(1-\sum_{1\leq q\leq R}\hat{\sigma}_{\ell\to i}^q)\right]\nonumber \\
\hat{\sigma}_{i\to j}^{r+1}&=&(1-p) \left\{\left[\prod_{\ell\in \partial i\setminus j}(1-\sum_{1\leq q\leq r-1}\hat{\sigma}_{\ell\to i}^q-\sum_{1\leq q\leq\bar{r}}\hat{\omega}_{\ell\to i}^q)-\prod_{\ell\in \partial i\setminus j}(1-\sum_{1\leq q\leq r}\hat{\sigma}_{\ell\to i}^q-\sum_{1\leq q\leq\bar{r}}\hat{\omega}_{\ell\to i}^q )\right]\right.\nonumber \\
&&+\theta({R-2r})\left[\prod_{\ell\in \partial i\setminus j}(1-\sum_{1\leq q\leq r}\hat{\sigma}_{\ell\to i}^q-\sum_{1\leq q\leq r-1}\hat{\omega}_{\ell\to i}^q)+\prod_{\ell\in \partial i\setminus j}(1-\sum_{1\leq q\leq  R-r}\hat{\sigma}_{\ell\to i}^q-\sum_{1\leq q\leq r}\hat{\omega}_{\ell\to i}^q)\right.\nonumber \\
&&\left.\left.-\prod_{\ell\in \partial i\setminus j}(1-\sum_{1\leq q\leq  r-1}\hat{\sigma}_{\ell\to i}^q-\sum_{1\leq q\leq r}\hat{\omega}_{\ell\to i}^q)-\prod_{\ell\in \partial i\setminus j}(1-\sum_{1\leq q\leq  R-r}\hat{\sigma}_{\ell\to i}^q-\sum_{1\leq q\leq r-1}\hat{\omega}_{\ell\to i}^q)\right]\right\}.
\label{eq:MP_sigma_hat}
\eea
Note that here and in the following we use $\bar{r}$ to indicate $\bar{r}=\min(r-1,R-r)$,  we use $\theta(x)$ to indicate the Heaviside function with $\theta(x)=1$ if $x>0$ and $\theta(x)=0$ otherwise, and we use $\partial i \setminus j$ to denote the neighborhood of node $i$ ($\partial i$) without node $j$.

The messages $\hat{\omega}_{i\to j}^r$   instead  obey
\bea
\hat{\omega}_{i\to j}^{1}&=&p\left[\prod_{\ell\in \partial i\setminus j}(1-\sum_{1\leq q\leq R}\hat{\sigma}_{\ell\to i}^q)\right],\nonumber \\
\hat{\omega}_{i\to j}^{r+1}&=&(1-p) \left[\prod_{\ell\in \partial i\setminus j}(1-\sum_{1\leq q \leq R-1}\hat{\sigma}_{\ell\to i}^r-\sum_{1\leq q\leq r-1}\hat{\omega}_{\ell\to i}^q)-\prod_{\ell\in \partial i\setminus j}(1-\sum_{1\leq q\leq R-1}\hat{\sigma}_{\ell\to i}^q-\sum_{1\leq q\leq r}\hat{\omega}_{\ell\to i}^q)\right].
\label{eq:MP_omega_hat}
\eea
\end{widetext}

The order parameter $P^{\infty}$ expressing the fraction of trusted nodes in the ERGC is given by
\bea
P^{\infty}=\frac{p}{N}\sum_{i=1}^N\left[1-\prod_{\ell\in \partial i}(1-\sum_{1\leq r\leq R}\hat{\sigma}_{\ell\to i}^r)\right].
\label{eq:P_infty_hat}
\eea
An untrusted node is in the ERGC if it is at distance less than $R$ from a trusted node in the ERGC. Thus we can define a second order parameter $U^{\infty}$ for ERP given by the fraction of untrusted nodes in the ERGC, expressed in terms of the messages as
\bea
U^{\infty}=\frac{(1-p)}{N}\sum_{i=1}^N\left[1-\prod_{\ell\in \partial i}(1-\sum_{1\leq r\leq R-1}\hat{\sigma}_{\ell\to i}^r)\right].
\label{eq:U_infty_hat}
\eea
 {The overall fraction of nodes in the ERGC is the sum between $P^{\infty}$ and $U^{\infty}$.}
Setting $R=1$, we recover the MP equations for standard percolation, \cite{bianconi2018multilayer}.

Note that the equations for the $\hat{\sigma}_{i \to j}^{r}$ in \eqref{eq:MP_sigma_hat} involve $R$ variables $\hat{\sigma}_{i \to j}^{r}$, with $1\leq r \leq R$, and only $\ceil{R/2}-1$ variables $\hat{\omega}_{i \to j}^r$,
with $1\leq r \leq \ceil{R/2}-1$.
Since Eqs. \eqref{eq:P_infty_hat} and \eqref{eq:U_infty_hat} for the order parameters contain only the variables $\hat{\sigma}_{i \to j}^{r}$, the total number of independent equations needed is $R+\ceil{R/2}-1$, rather than $2R$.
The MP equations \eqref{eq:MP_sigma_hat},\eqref{eq:MP_omega_hat}, and \eqref{eq:P_infty_hat},\eqref{eq:U_infty_hat}, can be used to compute the size of the ERGC on any simple network and for arbitrary interaction range $R$. They are exact in the limit of large $N$ for locally tree-like networks.
This theory can also be used as the starting point to derive the equations determining the critical properties of  ERP on random graphs with given degree distribution as explained in the following section.

\section{Theory for extended-range percolation on random graphs}
\subsection{General theoretical framework}
In this section we set up the general theoretical framework to study the critical properties of extended-range percolation on uncorrelated random graphs with arbitrary degree distribution $P(k)$.
For $R\leq 6$, these equations are equivalent to the equations derived in \cite{cirigliano2023extended}.
We stress, however, that the formalism in \cite{cirigliano2023extended} is rather cumbersome, and it is not easily  generalizable to arbitrary large values of $R$. Here, using the powerful MP theory developed above, we derive exact equations valid for arbitrary $R$.
The order parameters $P^{\infty}$ and $U^{\infty}$ determine the probability that a random node belongs to the ERGC and it is trusted or untrusted, respectively.
We can express these order parameters in terms of the probabilities $S_r$ that following a randomly chosen link we reach a node that is in the ERGC and is at distance $r-1$ from the closest trusted node, and of the probabilities $W_r$ that following a randomly chosen link we reach a node that is not in the ERGC and is at distance $r-1$ from the closest trusted node.

The probabilities $S_r$ and $W_r$ satisfy a set of self-consistent equations that can be derived from the MP equations discussed in the previous section.
Indeed, $S_r$ and $W_r$ can be identified as the average of the messages $\hat{\sigma}_{i\to j}^r$ and $\hat{\omega}_{i\to j}^r$, respectively, over the ensemble of random graphs with degree distribution $P(k)$ (see Appendix \ref{ApMP} for a more detailed discussion).

The equations for $S_r$ and $W_r$ read
\begin{widetext}
\bea
{S}_1&=&p\left[1-G_1\left(1-\sum_{1\leq q\leq R}S_q\right)\right],\nonumber \\
S_{r+1}&=&(1-p)\left\{\left[G_1\left(1-\sum_{1\leq q\leq r-1}S_q-\sum_{1\leq q\leq\bar{r}}W_q\right)-G_1\left(1-\sum_{1\leq q\leq r}S_q-\sum_{1\leq q\leq\bar{r}}W_q\right)\right]\right.\nonumber\\
&&+\theta(R-2r)\left[G_1\left(1-\sum_{1\leq q\leq r}S_q-\sum_{1\leq q\leq r-1}W_q\right)+G_1\left(1-\sum_{1\leq q\leq R-r}S_q-\sum_{1\leq q\leq r}W_q\right)\right.\nonumber \\
&&\left.\left.-G_1\left(1-\sum_{1\leq q\leq R-r}S_q-\sum_{1\leq q\leq r-1}W_q\right)-G_1\left(1-\sum_{1\leq q\leq r}S_q-\sum_{1\leq q\leq r}W_q\right)\right]\right\}\nonumber \\
{W}_1&=&pG_1\left(1-\sum_{1\leq q\leq R}S_q\right),\nonumber \\
W_{r+1}&=&(1-p)\left[G_1\left(1-\sum_{1\leq q\leq R-1}S_q-\sum_{1\leq q\leq r-1}W_q\right)-G_1\left(1-\sum_{1\leq q \leq  R-1}S_q-\sum_{1\leq q\leq r}W_q\right)\right],
\label{mes_ensemble}
\eea
\end{widetext}
where $G_0(x)$ and $G_1(x)$ are the generating functions defined as
\bea
G_0(x)=\sum_{k} P(k)x^k,\quad
G_1(x)=\sum_{k}\frac{kP(k)}{\avg{k}}x^{k-1}.
\eea

The probability  $P^{\infty}$ that a  node is trusted and is in the ERGC is given by
\bea
P^{\infty}=p\left[1-G_0\left(1-\sum_{1\leq r\leq R}S_r\right)\right],
\label{Pinf_c}
\eea
while the probability $U^{\infty}$ that a node is  untrusted and is in the is in the ERGC is given by
\bea
U^{\infty}=(1-p)\left[1-G_0\left(1-\sum_{1\leq r\leq R-1}S_r\right)\right].
\label{Uinf_c}
\eea
As noted in the previous section, the total number of independent equations needed is $R+\ceil{R/2}-1$, rather than $2R$, since the equations for the $S_r$ in \eqref{mes_ensemble} involve $R$ variables $S_r$ and only $\ceil{R/2}-1$ variables $W_r$.

In a more compact way, we can use the vector $\boldsymbol{Y}=(\boldsymbol{S}^{\top},\boldsymbol{W}^{\top})^{\top}$  encoding all the relevant $S_r$ ($\boldsymbol{S}=(S_1,S_2,\ldots)^{\top}$) and $W_r$ variables ($\boldsymbol{W}=(W_1,W_2,\ldots)^{\top}$) and we can write the equations \eqref{mes_ensemble} as
\bea
 \boldsymbol{Y} &= \boldsymbol{F}_{\boldsymbol{Y}}\big(\boldsymbol{Y},p\big)
 \label{ERP_compact}
 \eea
 or, alternatively, as
 \bea
  \boldsymbol{S} &= \boldsymbol{F}_{\boldsymbol{S}}\big(\boldsymbol{S},\boldsymbol{W},p),
   \label{ERP_compact_2a}\\
  \boldsymbol{W} &= \boldsymbol{F}_{\boldsymbol{W}}\big(\boldsymbol{S},\boldsymbol{W},p).
 \label{ERP_compact_2b}
 \eea

In Appendix \ref{AppendixA} we write down explicitly the equations \eqref{mes_ensemble} and \eqref{Pinf_c} for $R \leq 4$. These equations are shwon to be equivalent to the ones derived in \cite{cirigliano2023extended} in Appendix \ref{AppendixERP}.
\begin{figure}[htb!]
   \centering
   \includegraphics[width=\columnwidth]{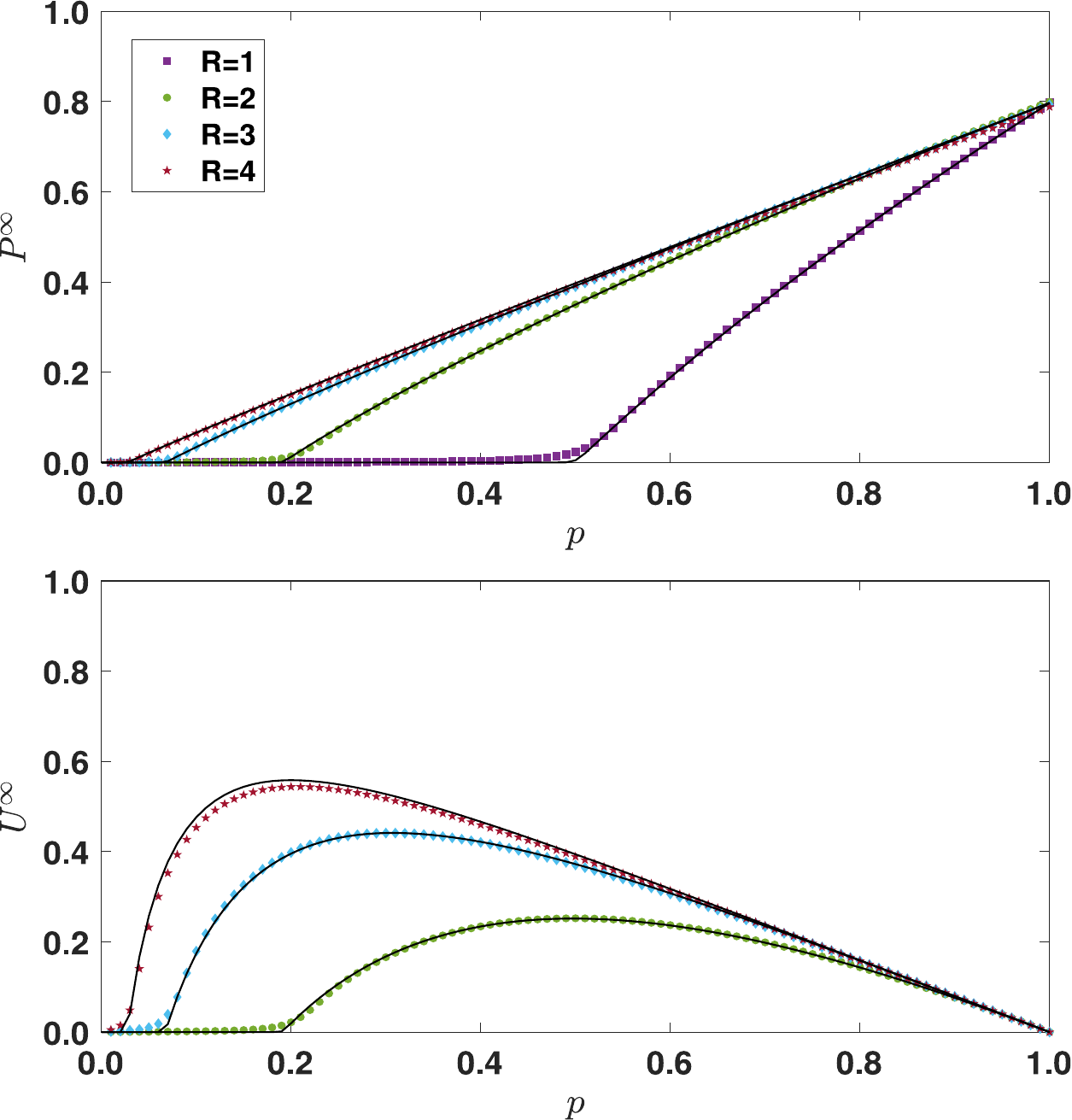}
   \caption{The ERP order parameters, $P^{\infty}$ and $U^{\infty}$, as functions of the fraction of trusted nodes $p$ for $R \in {1,2,3,4}$.
   ERP exhibits a continuous phase transition at the percolation threshold $p=p_c$, which decreases as $R$ increases. Note that for $R=1$, $P^{\infty}$ reduces to standard percolation and that  $U^{\infty}$ is only non-trivial (i.e. non-zero) for $R>1$. The symbols indicate results of Monte Carlo simulations over $1000$ realizations of Poisson networks of $N=2\times 10^4$ nodes and average degree $c=2$.}
   \label{fig:order_parameters_ERP}
\end{figure}
\subsection{Phase transition and critical threshold}
The ERP order parameters $P^{\infty}$ and $U^{\infty}$ as functions of
$p$ are plotted in Fig.~\ref{fig:order_parameters_ERP} for a Poisson network of average
degree $c=2$. The analytic solution is in perfect agreement with
results of numerical simulations.  Extended-range percolation displays
a continuous phase transition at $p=p_c$, in analogy with standard
percolation (case $R=1$).  As intuitively expected, the percolation
threshold decreases as $R$ increases: the presence of a range $R>1$
enhances the robustness of the network. Moreover, while the order
parameter $P^{\infty}$ is monotonically increasing in $p$, the order
parameter $U^{\infty}$ displays a maximum as a function of $p$.  The
physical intuition is as follows.  For $p<p_c$ both $P^{\infty}$ and
$U^{\infty}$ are identically zero as there are no nodes in the
ERGC. For $p=1$, the ERGC coincides with the giant component of the
network for any $R$, so $P^{\infty}(R)=P^{\infty}$ while
$U^{\infty}(R)=0$.  Thus $U^{\infty}$ cannot be monotonically
increasing and actually displays a single maximum. 

As anticipated before, extended-range percolation is characterized by
a continuous phase transition at the percolation threshold
$p_c$, where $P^{\infty}$ becomes non-zero. For $p \leq p_c$,
there is a unique solution for Eqs.~\eqref{ERP_compact},
$S_r=S_r^{0}=0$ for all values of $r$ and $W_r=W^{0}_r$ with
\bea
W^{\star}_1&=&p,\nonumber
\\ W_{r+1}^{\star}&=&(1-p)\left[G_1\Big(1-\sum_{1\leq q\leq
    r-1}W_{q}^{\star}\Big)\right.\nonumber
  \\&&\left.-G_1\Big(1-\sum_{1\leq q\leq r}W_{q}^{\star}\Big)\right],
\eea
corresponding to $P^{\infty}=U^{\infty}=0$. For $p>p_c$ another
solution with positive $S_r$ appears, continuously in $p-p_c$. Let us
denote $\boldsymbol{Y}^{0}=({\boldsymbol{S}^{0
    \top}},\boldsymbol{W}^{0\top})^{\top}$.
The percolation threshold $p_c$ can be obtained linearizing Eq.~\eqref{ERP_compact_2a}
around $\boldsymbol{S}=\boldsymbol{S}^{0}$ and imposing that
\bea
\Lambda_0(p_c)=1,
\eea
where $\Lambda_0$ is the largest
eigenvalue of the
Jacobian matrix of $\boldsymbol{F}_{\boldsymbol{Y}}$ performed with
respect to the variables in $\boldsymbol{Y}$ and calculated in
$\boldsymbol{Y}^{0}$.  As shown in
\cite{cirigliano2023extended}, ERP on
uncorrelated random graphs belongs to the same universality class of
standard percolation for homogeneous degree distribution and for
power-law degree distributions $P(K) \sim k^{-\gamma}$ with $\gamma >
3$, while $R$-dependent critical exponents are found for strongly
heterogeneous networks with $2 < \gamma <3$.

\section{Multiplex extended-range percolation}
\subsection{Theoretical framework}
Consider a multiplex network \cite{bianconi2018multilayer} $\vec{\mathcal{G}}=(\mathcal{G}^{[1]},\mathcal{G}^{[2]},\ldots,\mathcal{G}^{[\alpha]},\ldots, \mathcal{G}^{[M]})$ formed by $M$ network layers $\mathcal{G}^{[\alpha]}=(V,E^{[\alpha]})$ of $N=|V|$ nodes, with $\alpha\in \{1,\ldots, M\}$. Each network $\mathcal{G}^{[\alpha]}$ is drawn independently at random from the ensemble of uncorrelated random networks with  degree distribution $P^{[\alpha]}(k)$, whose generating functions are
\bea
G_0^{[\alpha]}(x)&=&\sum_{k} P^{[\alpha]}(k)x^k,\nonumber \\
G_1^{[\alpha]}(x)&=&\sum_{k}\frac{kP^{[\alpha]}(k)}{\avg{k}}x^{k-1}.
\eea
We assume that the multiplex network has thus a negligible link overlap \cite{bianconi2013statistical}. Each node of the multiplex network is either a trusted node on every layer or an untrusted node on every layer.

\begin{figure}[!htb]
   \centering
   \includegraphics[width=0.95\columnwidth]{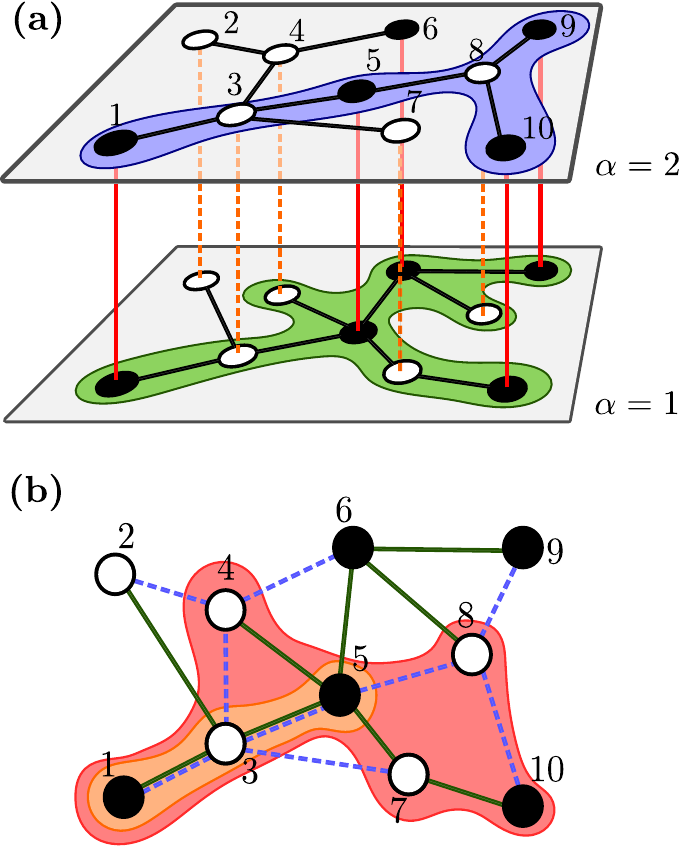}
   \caption{Schematic representation of multiplex extended-range
     percolation for $R=2$ and $M=2$ layers. Trusted and untrusted
     nodes are indicated by filled and empty circles,
     respectively. (a) The ERGC of each single layer is highlighted,
     in green for layer $\alpha=1$ and in blue for layer
     $\alpha=2$. Red solid interlayer links represent the interdependencies
     for trusted nodes in both versions of MERP, yellow
     dashed interlayer links represent the condition on untrusted
     nodes, considered only in version 1. Note that some nodes are
     facilitators in only one layer, e.g. node $7$, while others are
     facilitators in both layers, e.g. node $3$. (b) The two,
     different, MERGC for version 1, highlighted in yellow, and
     version 2 highlighted in red. Since the trusted node $6$ is not
     in the ERGC in layer $2$, it cannot be part of the MERGC, causing
     also node $9$ to disconnect. The untrusted node $7$ is in the
     ERGC of layer $1$, but not in layer $2$. As a consequence, it is
     in the MERGC for version $2$ but not for version $1$, and it also
     affects the presence in the MERGC of node $10$. The size of the MERGC for
     the two versions of the model can thus be significantly
     different.}
   \label{fig:schematic_MERP}
\end{figure}

Inspired by the multiplex interdependent percolation (MIP) model \cite{buldyrev2010catastrophic,
baxter2012avalanche,bianconi2018multilayer},
we formulate two versions of the Multiplex Extended-Range Percolation (MERP) model: version $V=1$ and version $V=2$ (see Fig. \ref{fig:schematic_MERP} for a schematic representation of the models). Both versions exploit the presence of untrusted facilitator nodes, and the extended-range percolation mechanism together with the interdependencies for nodes of different layers.

The Multiplex  Extended-Range Giant Component
(MERGC) is defined as follows.  In both versions of the model, a
trusted node is in the MERGC if it belongs to the ERGC {\em in each
  layer} of the multiplex network. Version 1 and version 2 impose
however different conditions on untrusted nodes. In Version 1, we
require that each untrusted node in the MERGC must be in the ERGC {\em
  in each layer} of the multiplex, as it is required for trusted
nodes. In Version 2 instead, untrusted nodes in the MERGC must be part
of the ERGC in {\em at least} one layer. This difference in the
condition imposed on untrusted nodes strongly affects the
robustness of the multiplex, as explained in detail below. Let us make
some qualitative preliminary remarks on the behaviour of MERP. As the ERGC
on a single network reduces, for any finite $R$, to the
standard giant component in the network for $p=1$ (untrusted nodes are
absent), in both versions of MERP the MERGC reduces for $p=1$ to the
mutually connected giant component of MIP. For $R=1$ untrusted nodes
have no role, thus both versions of MERP have the same MERGC, which
coincides with the mutually connected giant component (MCGC) of MIP.

The exact solution of MERP on uncorrelated random graphs with
arbitrary degree distribution can be obtained as follows. First, we
introduce the variables $S_{r}^{[\alpha]}$ and $W_{r}^{[\alpha]}$,
defined as the probabilities $S_r$ and $W_r$ in layer $\alpha$. Then,
we can define recursive equations for these variables starting from
Eqs.~\eqref{mes_ensemble}, \eqref{Pinf_c}, \eqref{Uinf_c}, and
implementing the different conditions required by the two versions of
MERP. The requirement for trusted nodes to be part of the ERGC in all
layers can be easily implemented with the multiplicative factor
\[ \prod_{\beta \neq \alpha}\left[1-G_0^{[\beta]}\left(1-\sum_{1\leq r\leq R}S_r^{[\beta]}\right)\right] \]
in the equations for $S_{1}^{[\alpha]}$ and $W_{1}^{[\alpha]}$, thus obtaining the equations
\bea
{S}_1^{[\alpha]}&=&p\left[1-G_1^{[\alpha]}\left(1-\sum_{1\leq r\leq R}S_r^{[\alpha]}\right)\right]\nonumber \\&&\left\{\prod_{\beta\neq \alpha}\left[1-G_0^{[\beta]}\left(1-\sum_{1\leq r\leq R}S_r^{[\beta]}\right)\right]\right\},\nonumber \\
{W}_1^{[\alpha]}&=&p\left[G_1^{[\alpha]}\left(1-\sum_{1\leq r\leq R}S_r^{[\alpha]}\right)\right]\nonumber \\
&&\left\{\prod_{\beta\neq \alpha}\left[1-G_0^{[\beta]}\left(1-\sum_{1\leq r\leq R}S_r^{[\beta]}\right)\right]\right\}.
\eea
The requirement for untrusted nodes to be in the ERGC in all layers can be instead implemented by means of
the multiplicative factor
\[\prod_{\beta \neq \alpha}\left[1-G_0^{[\beta]}\left(1-\sum_{1\leq r\leq R-1}S_r^{[\beta]}\right)\right] \]
in the equations for $S_{r}^{[\alpha]}$ and $W_{r}^{[\alpha]}$ with $2\leq r \leq R$. However, as explained above, this condition is required only in version $V=1$ of the model, while no restriction on untrusted nodes is applied for $V=2$. The two different conditions for untrusted nodes can then be implemented all at once using Kronecker's $\delta_{V,1}$ and $\delta_{V,2}$. We get the equations
\begin{widetext}
\bea
S_{r+1}^{[\alpha]}&=&(1-p)\left\{\left[G_1^{[\alpha]}\left(1-\sum_{1\leq q\leq r-1}S_q^{[\alpha]}-\sum_{1\leq q\leq\bar{r}}W_q^{[\alpha]}\right)-G_1^{[\alpha]}\left(1-\sum_{1\leq q\leq r}S_q^{[\alpha]}-\sum_{1\leq q\leq\bar{r}}W_q^{[\alpha]}\right)\right]\right.\nonumber\\
&&+\theta(R-2r)\left[G_1^{[\alpha]}\left(1-\sum_{1\leq q\leq r}S_q^{[\alpha]}-\sum_{1\leq q\leq r-1}W_q^{[\alpha]}\right)+G_1^{[\alpha]}\left(1-\sum_{1\leq q\leq R-r}S_q^{[\alpha]}-\sum_{1\leq q\leq r}W_q^{[\alpha]}\right)\right.\nonumber \\
&&\left.\left.-G_1^{[\alpha]}\left(1-\sum_{1\leq q\leq R-r}S_q^{[\alpha]}-\sum_{1\leq q\leq r-1}W_q^{[\alpha]}\right)-G_1^{[\alpha]}\left(1-\sum_{1\leq q\leq r}S_q^{[\alpha]}-\sum_{1\leq q\leq r}W_q^{[\alpha]}\right)\right]\right\}\nonumber \\
&&\times\left\{\delta_{V,2}+\delta_{V,1}\prod_{\beta\neq \alpha}\left[1-G_0^{[\beta]}\left(1-\sum_{1\leq r\leq R-1}S_r^{[\beta]}\right)\right]\right\},\nonumber \\
W_{r+1}^{[\alpha]}&=&(1-p)\left[G_1^{[\alpha]}\left(1-\sum_{1\leq q \leq R-1}S_q^{[\alpha]}-\sum_{1\leq q\leq r-1}W_q^{[\alpha]}\right)-G_1^{[\alpha]}\left(1-\sum_{1\leq q \leq R-1}S_q^{[\alpha]}-\sum_{1\leq q\leq r}W_q^{[\alpha]}\right)\right]\nonumber\\
&&\times\left\{\delta_{V,2}+{\delta_{V,1}\prod_{\beta\neq \alpha}\left[1-G_0^{[\beta]}\left(1-\sum_{1\leq r\leq R-1}S_r^{[\beta]}\right)\right]}\right\}.
\label{mes}
\eea
\end{widetext}

Using a notation similar to the one adopted for ERP, we can write these equations in a more compact way. Introducing the vector $\boldsymbol{Y}=(\boldsymbol{S}^{\top},\boldsymbol{W}^{\top})^{\top}$ and encoding all the relevant $S_r^{[\alpha]}$ ($\boldsymbol{S}=(S_1^{[1]},S_2^{[1]},\ldots S^{[\alpha]} \ldots )^{\top}$) and $W_r^{[\alpha]}$ variables ( $\boldsymbol{W}=(W_1^{[1]},W_2^{[2]},\ldots,W^{[\alpha]}_r \ldots)^{\top}$), we can write the  equations \eqref{mes}, as
\bea
 \boldsymbol{Y} &= \boldsymbol{F}\big(\boldsymbol{Y},p).
 \label{MERP_compact}
 \eea

For the order parameter $P^{\infty}$, corresponding to the probability that a random node is in the MERGC and it is trusted, we can write the equation
\bea
P^{\infty}=p\prod_{\alpha=1}^M\left[1-G_0^{[\alpha]}\left(1-\sum_{1\leq r\leq R}S_r^{[\alpha]}\right)\right],
\label{eq:P_inf_MERP}
\eea
valid for both versions of MERP. For the probability $U^{\infty}$ that a node is untrusted and belongs to the  MERGC we must distinguish between $V=1$ and $V=2$. In version 1, we require that untrusted nodes in the MERGC must be in the ERGC in all layers, hence we have
\bea
U^{\infty}=(1-p)\prod_{\alpha=1}^M\left[1-G_0^{[\alpha]}\left(1-\sum_{1\leq r\leq R-1}S_r^{[\alpha]}\right)\right].
\label{eq:U_inf_MERP1}
\eea
In version 2 we require that an untrusted node in the MERGC is in the ERGC in at least one layer, hence we get
\bea
U^{\infty}=(1-p)\left[1-\prod_{\alpha=1}^MG_0^{[\alpha]}\left(1-\sum_{1\leq r\leq R-1}S_r^{[\alpha]}\right)\right].
\label{eq:U_inf_MERP2}
\eea
\begin{figure*}[htb!]
   \centering
   \includegraphics[width=0.95\textwidth]{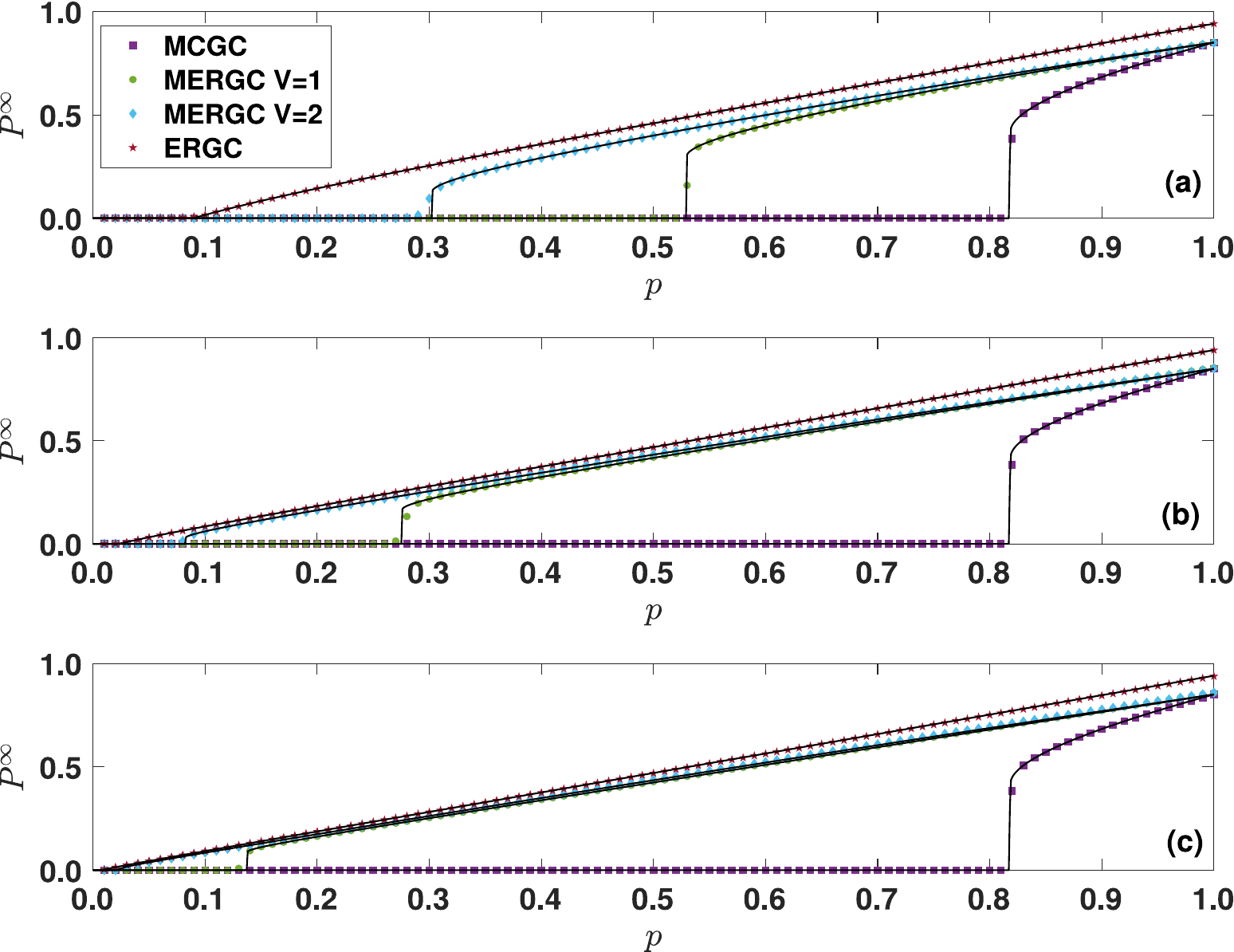}
   \caption{Comparison between the order parameter $P^{\infty}$ as a function of $p$ for the MIP, the two versions of MERP (MERP $V=1$ and MERP $V=2$),
     and the single layer ERP, for (a) $R=2$, (b) $R=3$ and (c) $R=4$.
     The results of the Monte Carlo simulations (symbols) are in
     excellent agreement with our theoretical predictions (solid
     lines). The considered multiplex networks have $M=2$ uncorrelated layers and $N=10^5$ nodes. Each layer (or single network for ERP) is formed by a Poisson network with average degree $c=3$. Data are averaged over $20$ runs.}
   \label{fig:MERGC_a}
\end{figure*}

As it will be discussed in detail for a particular case in the next
section, MERP may display one or more discontinuous hybrid
transitions. The critical value $p=p^{\star}$ for these transitions
can be in general found imposing that Eqs.~\eqref{MERP_compact} admits
a nontrivial solution
$\boldsymbol{Y}=\boldsymbol{Y}^{\star}=(\boldsymbol{S}^{\star},\boldsymbol{W}^{\star})$
with $\boldsymbol{S}^{\star}\neq {\bf 0}$, and that the maximum
eigenvalue $\Lambda_{\star}$ of the Jacobian of the function
$\boldsymbol{F}$ with respect to $\boldsymbol{Y}$ evaluated at
$\boldsymbol{Y}=\boldsymbol{Y}^{\star}$ is one, i.e. $p^{\star}$ is
the solution of the equations
\bea \boldsymbol{Y}^{\star} &=
\boldsymbol{F}\big(\boldsymbol{Y}^{\star},p^{\star}),\quad
\Lambda_{\star}(p^{\star})=1.
\eea
Note the difference with the
criticality condition of ERP, and in general of systems exhibiting
continuous phase transitions. While for a continuous transition the
criticality condition is determined by the largest eigenvalue of the
Jacobian evaluated at the trivial fixed point $\boldsymbol{Y}^{0}$,
for discontinuous transitions the critical threshold is determined
when the largest eigenvalue of the Jacobian evaluated in
$\boldsymbol{Y}^{\star}$, the nontrivial solution of
Eqs.~\eqref{MERP_compact}, equals 1.

\begin{figure*}[htb!]
   \includegraphics[width=0.95\textwidth]{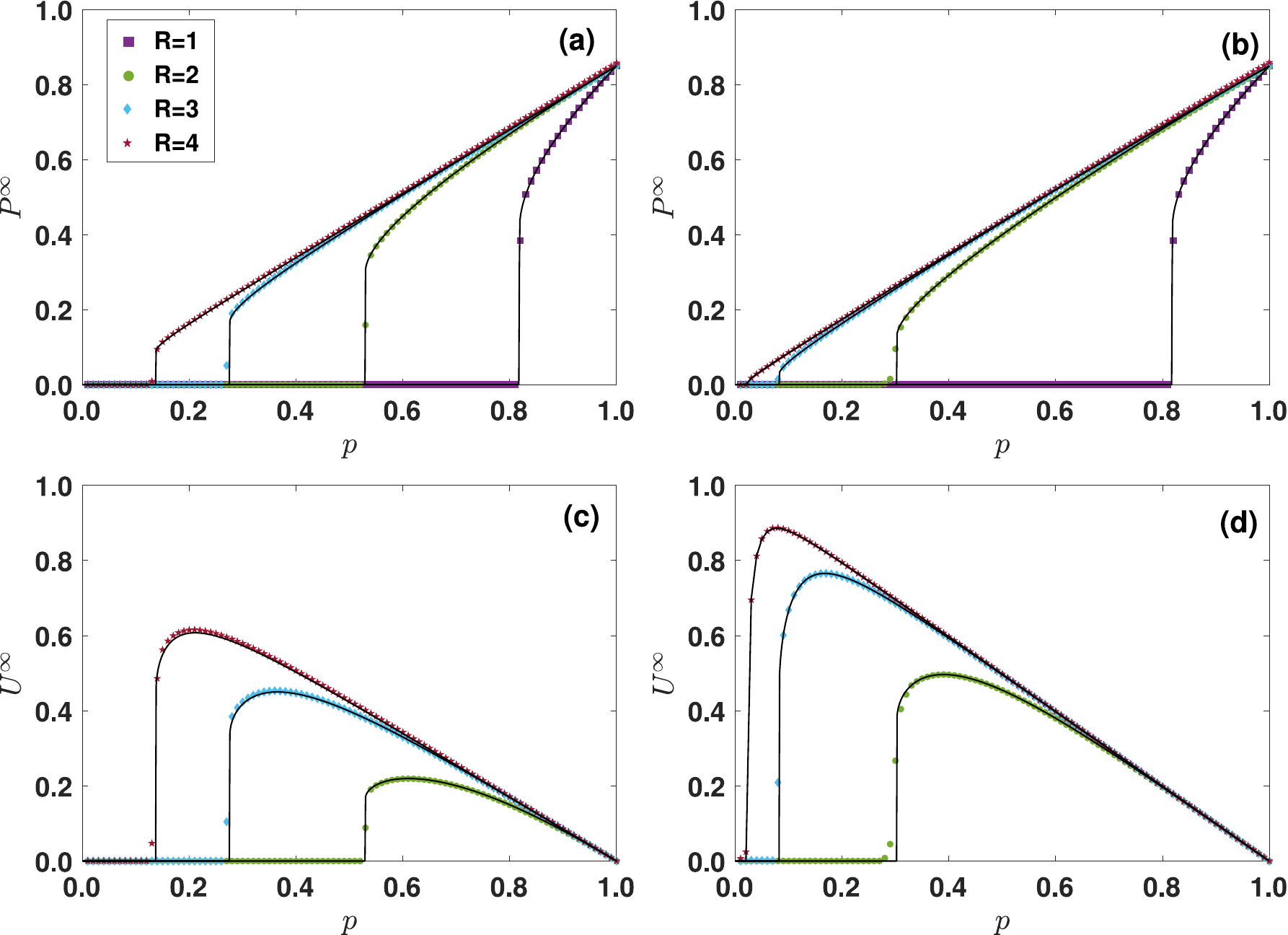}
   \caption{The order parameters $P^{\infty}$ and $U^{\infty}$ of  MERP for version $V=1$ (panels (a) and (c)) and version $V=2$ (panels (b) and (d)) are plotted as a function of $p$ for different values of $R\in \{1,2,3,4\}$ (note that for $R=1$ the result reduce to the standard mutually connected giant component-MCGC). The discontinuous (hybrid) transition observed in Monte Carlo simulations (symbols) is in excellent agreement with our theoretical predictions (solid lines).
   The considered multiplex networks have $M=2$ uncorrelated layers and $N=10^5$ nodes. Each layer is formed by a Poisson network with average degree $c=3$. Data are averaged over $20$ runs.}
   \label{fig:MERGC_b}
\end{figure*}

\begin{figure*}[htb!]
   \includegraphics[width=0.95\textwidth]{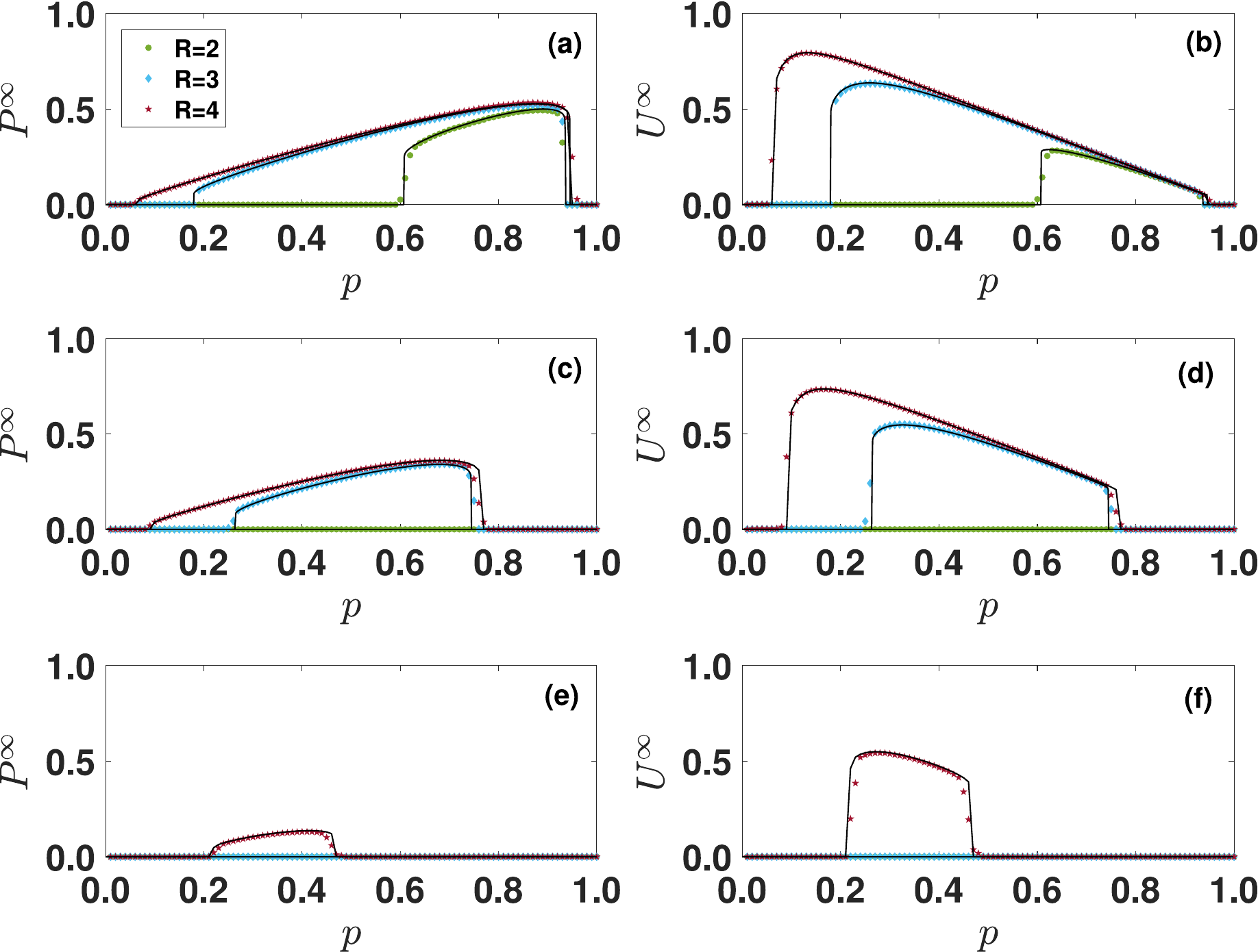}
\caption{The order parameters $P^{\infty},U^{\infty}$ for MERP, version $V=2$ and $R\in \{2,3,4\}$ are shown to display a reentrant phase transition for multiplex networks that do not display a MCGC.
The considered multiplex networks have $M=2$ uncorrelated layers and $N=10^5$ nodes. Each layer is formed by a Poisson network with average degree $c=2.4$ (panels (a) and (b)), $c=2.2$ (panels (c) and (d)) and $c=1.9$ (panels (e) and (f)). As the average degree $c$ decrease, the MERGC disappears for MERP of increasing values of $R$.
Data are averaged over $20$ runs. The results of the Monte Carlo simulations (symbols)  are in excellent agreement with our theoretical predictions (solid lines).}
   \label{fig:MERGC_reentrant}
\end{figure*}

\subsection{Phenomenology of the  MERP for $R\in \{2,3,4\}$ on Poisson multiplex networks}

In this section, we study the critical properties of MERP for $R\in \{2,3,4\}$ on
multiplex networks composed of $M$ layers independently drawn from the random Poisson networks ensemble\footnote{Poisson networks are Erd\H{o}s-R\'enyi random graphs with link probability $p=c/(N-1)$ in the limit $N \gg 1$, which define random graphs with Poisson degree distribution $P(k)=e^{-c}c^k/k!$.} with average degree $c$ (where $c$ is the same for each layer).
Specifically, here we focus on the relevant phenomenology predicted by our theoretical approach and we test it by Monte Carlo simulations of the process on multiplex networks with $M=2$ layers. We defer a deeper theoretical derivation of the phase diagram to the next subsection.

As a general remark, we expect that a non-zero MERGC for version $V=2$
will be easier to achieve than for version $V=1$, because version
$V=2$ does not require interdependence of the untrusted facilitator
nodes. Thus the MERGC of version $V=2$ is always expected to be larger
or at most equal to the MERGC of version $V=2$ for the same choice of
the parameters. As we have already remarked the MERGC for $R>1$ are
also expected to be larger than or equal to the MCGC, which is equivalent
to the MERGC for $R=1$. In particular both versions of the MERGC
coincide with the MCGC for $p=1$, independently of the value of $p$.
At the same time, if the layers are independently drawn random
networks with the same degree distribution, the MERGC of both versions
of the model will always be smaller than the ERGC of a single
layer. For an illustration of the relation between the MERGC, the ERGC
and the MCGC see Figure~\ref{fig:MERGC_a}.

Let us indicate with $\bar{c}$ the minimum average degree required to
observe a MCGC (or equivalently a MERGC with $R=1$).
Its value for $M=2$ is $\bar{c}=2.45541\ldots$.
For version $V=1$ of MERP $\bar{c}$
also indicates the minimum average degree required to observe a MERGC
independently of the value of $R\geq 1$. In version $V=1$ and as long as $c>\bar{c}$ also in version $V=2$ we observe a single hybrid discontinuous phase transition at the emergence of the MERGC (see Fig.~\ref{fig:MERGC_b}).

For version $V=2$ of MERP we observe a highly nontrivial
behavior for average degree $c$ in an interval of values $c^{\star}<c< \bar{c}$ where  $c^{\star}$ decreases if $R$ increases (see Figure $\ref{fig:MERGC_reentrant}$).
For small $p$ there is no giant component. Increasing $p$, at some point
$p^{\star}_{-}$ the usual discontinuous transition for percolation on multiplex
networks takes place, after which the size of the MERGC increases.
However, the behavior is nonmonotonic: $P^{\infty}$ reaches
a maximum and for a value $p_{+}^{\star}<1$ the MERGC discontinuously disappears.
As we discuss in the next section, both transitions at $p_{-}^{\star}$ and $p_{+}^{\star}$ are hybrid. Thus our results demontrate the existence of a re-entrant discontinuous transition for version $V=2$ of MERP.

The physical interpretation of this phenomenology is as follows.
In the limit of $p$ close to 1, almost all nodes are trusted,
untrusted nodes do not really play a role and version $V=2$ is practically
equivalent to version $V=1$, which does not admit a MERGC for $c<\bar{c}$.
Hence $P^{\infty}=0$ for $p \to 1$ for both versions, and for any $R$.
When $p$ is reduced from 1 the fraction of untrusted nodes increases;
their presence facilitates the formation of the MERGC for $V=2$,
since they can play the role of untrusted facilitator nodes,
each in its layer, with no interdependencies. This leads to the discontinuous
formation, at $p_{+}^{\star}$, of the MERGC.
Clearly, when $p$ is reduced further this positive effect starts to be
offset by the decrease of trusted nodes, leading to the breakdown of the
MERGC for $p_{-}^{\star}$.
Hence there is a finite interval $[p_{-}^{\star},p_{+}^{\star}]$ in which the presence of many
but not too many noninterdependent untrusted nodes allows the existence of a MERGC.

\subsection{Phase diagram of MERP for $R=2$ on Poisson multiplex networks}

After recalling the main
results~\cite{son2012percolation,buldyrev2010catastrophic,bianconi2018multilayer} valid for the
MIP (i.e. the MERP with $R=1$), in this subsection we derive the phase
diagram of MERP for $R=2$ on Poisson multiplex networks with $M$ layers
having the same average degree $c$. A similar argument holds for
MERP with $R>2$ on the same multiplex networks.

The fact that layers have identical statistical properties
significantly simplifies the equations for MERP.
Indeed, we can safely assume that $S^{\alpha}_r=S_r$ and $W^{\alpha}_r=W_r$
do not depend on $\alpha$.
Moreover, since the layers of the multiplex network have a
Poisson degree distribution, $G_0(x)=G_1(x)$, hence $P^{\infty}=S_1$.

\subsubsection{Case $R=1$ (MCGC)}
Let us briefly recall the main results ~\cite{son2012percolation,buldyrev2010catastrophic,bianconi2018multilayer}  valid for $R=1$,
where the MERGC reduces to the mutually connected giant component MCGC
of the multiplex network.
For $R=1$ we have that $S_1=P^{\infty}$ obeys
\bea
S_1=F_1^{(0)}(S_1) \equiv p(1-e^{-cS_1})^M.
\eea
Thus, for each value of $p$, $c$  and $M$ the fraction $S_1$ of nodes in the MCGC can be found by considering the zeros of the function
\bea
\tilde{H}^{(0)}(S_1) \equiv S_1-F_1^{(0)}(S_1)=S_1-p(1-e^{-cS_1})^M.
\eea
Note that for each choice of the parameters only the largest stable solution is physical.
Interestingly, in this case we can further reduce the parameters by introducing the auxiliary variables $x=S_1/p$, $z=cp$, and studying the zeros of the function
\bea
H^{(0)}(x,z)=x-(1-e^{z x})^M.
\eea
For any fixed value of $M$ we can thus study the zeros of  $H^{(0)}(x)$
as a function of the product $z=cp$.
In this way it is found that for every $M>1$  the MCGC emerges at a
discontinuous (hybrid) transition where $x=x^{\star}$ and $z=z^{\star}$ are
determined by the equations
\bea
H^{(0)}(x^{\star},z^{\star})=0,\quad \left.\frac{\partial H^{(0)}}{\partial x}\right|_{x^{\star},z^{\star}}=0.
\eea
The value of $z^{\star}=\bar{c}$ determining the minimal average degree $\bar{c}$ for observing the MCGC, increases as a function of $M$. For $M=2$ we obtain in this way $\bar{c}=2.45541\dots$ .
\begin{figure*}[!htb!]
   \includegraphics[width=\textwidth]{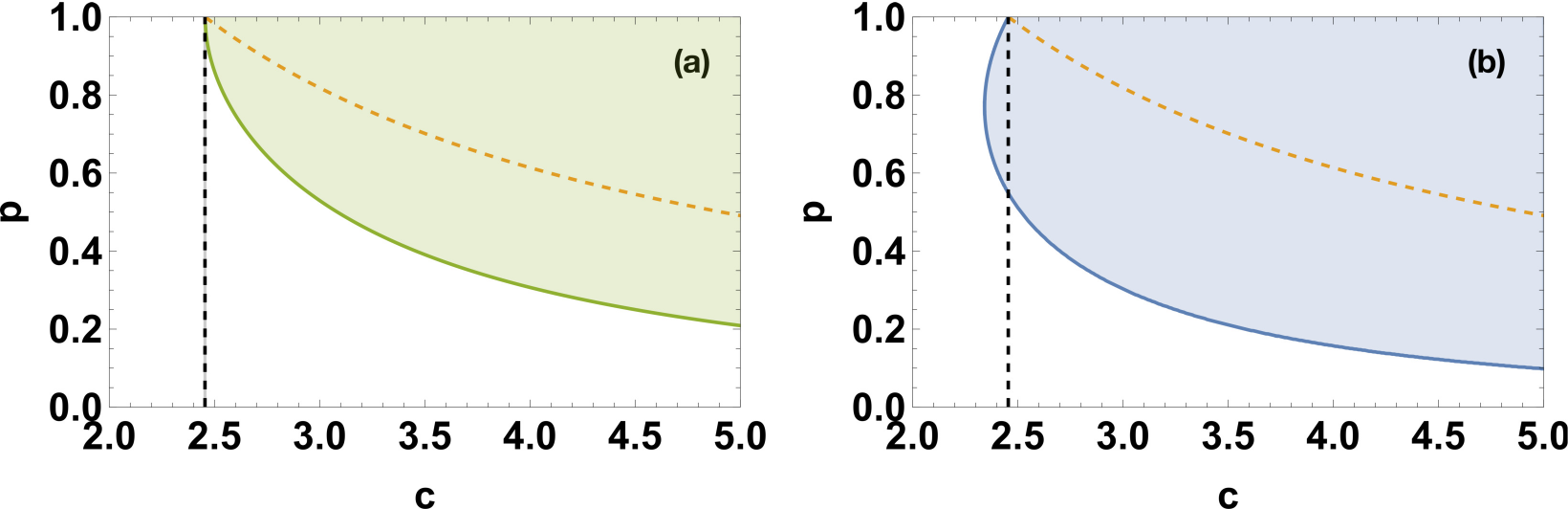}
   \caption{Phase diagram of $R = 2$ MERP on Poisson multiplex networks with $M = 2$ layers and average degree $c$. Panel (a) and (b)
display the phase diagram for version $V = 1$ and $V = 2$ of MERP, respectively. The filled regions are the region corresponding to a non-zero MERGC. In both panels, the orange dashed line correspond to the critical line for $R = 1$ MERP, i.e. the critical line for the emergence of the MCGC; the vertical dashed line $c = \bar{c} = 2.455$ indicates the critical average degree for the emergence of the MCGC. We observe that for version $V = 2$ there are values $c^{\star}<c<\bar{c}$ in which the reentrant phase transition is predicted, providing solid understanding of the Monte Carlo results.}
   \label{fig:phase_diagram}
\end{figure*}
\subsubsection{Case $R=2$ }

We now move to the interesting MERP case with $R=2$.  In the
simplified setting of multiplex Poisson networks, the equations for
version $V$ read
\bea
S_1&=&F_1^{(V)}(S_1,S_2)=p\left(1-e^{-c(S_1+S_2)}\right)^M,\nonumber
\\ S_2&=&F_2^{(V)}(S_1)=(1-p)\left(1-e^{-cS_1}\right)^{M
  \delta_{V,1}+\delta_{V,2}}, \eea where again we use the Kronecker
delta to implement the two versions $V=1$ and $V=2$ all at once.  As a
first remark, note that for $p=1$ these equations coincide with the
equations for MCGC for values of $V$, see
Fig.~\ref{fig:MERGC_a}. Indeed for $p=1$ all nodes are trusted,
hence there is no difference between the MERGC and the MCGC, as
already noted above. For $p < 1$, instead the MERGC strongly depends
on the version considered and is different from the MCGC.

Since for both versions $F_2^{(V)}$ is a function of $S_1$ only, in
order to investigate the critical properties of the two models we
can simply study the solutions of the equation \bea
S_1=F_1^{(V)}\Big(S_1,F_2^{(V)}(S_1)\Big), \eea or equivalently the
zeros of the function
\bea
H^{(V)}(S_1;c,p)=S_1-F_1^{(V)}\Big(S_1,F_2^{(V)}(S_1)\Big),
\label{eq:H_poisson}
\eea
for $V=1$ and $V=2$. Note however that this function depends
independently from $c$ and $p$, thus we do not have in general the
same simplification that we discussed for MCGC. Studying
Eq.~\eqref{eq:H_poisson} one finds that for any $M>1$ and for any $c$
the percolation threshold $p=p^{\star}$ at which the MERGC emerges at a
discontinuous (hybrid) transition where the order parameter
$P^{\infty}(p_\star)=S_1^{\star}$ can be determined by solving the
equations \bea H^{(V)}(S_1^{\star};p_\star,c)=0,\quad
\left.\frac{\partial H^{(V)}}{\partial S_1}\right|_{S_1^{\star},p_\star}=0.
\label{dis}
\eea
To show that these transitions are hybrid we can study the solution of $H^{(V)}(S_1;p,c)=0$ for $\delta p=p-p_\star \ll 1$  and $\delta  S_1 = S_1-S_1^{\star}\ll 1$.
Thus expanding $H^{(V)}(S_1;p,c)$ and using Eq.\eqref{dis}, we obtain
\bea
0&=&H^{(V)}(S_1;p,c)\simeq \left.\frac{\partial  H^{(V)}}{\partial p}\right|_{S_1^{\star},p_\star}\delta p\nonumber\\
&&+\frac{1}{2}\left.\frac{\partial^2 H^{(V)}}{\partial S_1^2}\right|_{S_1^{\star},p^{\star}}(\delta S_1)^2.
\eea
It follows that if
\bea
\left.\frac{\partial^2 H^{(V)}}{\partial S_1^2}\right|_{S_1^{\star},p^{\star}}>0,\quad \left.\frac{\partial H^{(V)}}{\partial p}\right|_{S_1^{\star},p_\star}<0,
\eea
the transition is hybrid with a square root singularity, i.e.,
\bea
\delta S_1 \propto (\delta p)^{\beta}
\eea
with $\beta=1/2.$

For version 1 of MERP, Eqs.~\eqref{dis} have a single non-trivial
solution as long as $c>{c}^{\star}$ where $c^{\star}=\bar{c}$, i.e., we
have a nonvanishing MERGC only for average degrees such that also the MCGC
is non-zero.  For version 2 of MERP instead, in a certain range of $c$
values, we observe two discontinuous hybrid transitions, both
satisfying Eq.~\eqref{dis}, one determining the onset of the MERGC and
one determining its dismantling, see Fig.~\ref{fig:MERGC_reentrant}. This occurs for average degrees
$c \in [c^{\star},\bar{c}]$ where $c^{\star}$ can be found imposing the
equations \bea H^{(2)}(S_1^{\star};p^{\star},c^{\star})=0,\nonumber
\\ \left.\frac{\partial H^{(2)}}{\partial
  S_1}\right|_{S_1^{\star},p^{\star},c^{\star}}=0,\nonumber
\\ \left.\frac{\partial H^{(2)}}{\partial p}\right|_{S_1^{\star},p^{\star},c^{\star}}=0.
\label{dis2}
\eea

\begin{figure}[htb!]
   \includegraphics[width=0.98\columnwidth]{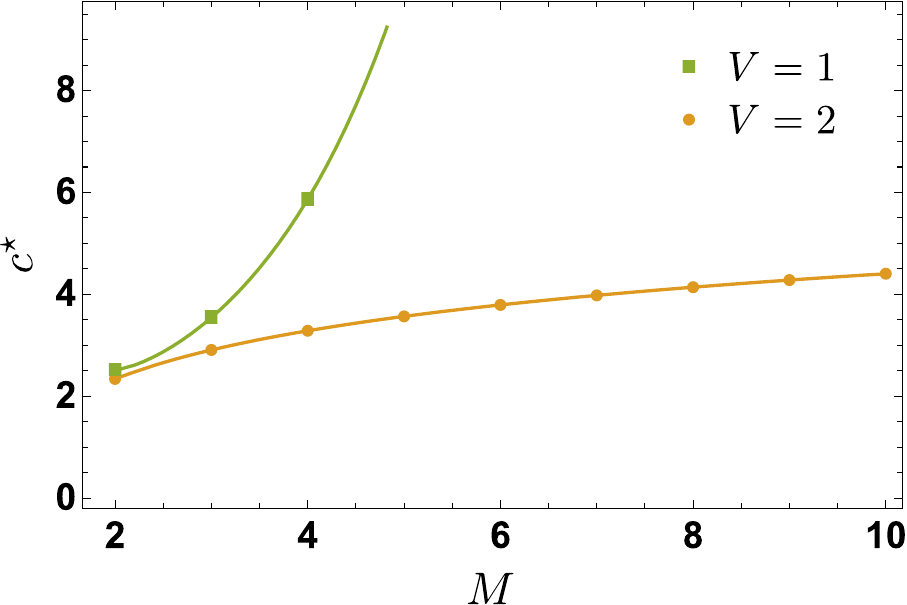}
   \caption{The critical average degree $c^{\star}$ of a Poisson
     multiplex network which allows the emergence of a $R=2$ MERGC is
     plotted versus the number of layers $M$ for the version $V=1$
     (green, square symbols) and for version $V=2$ (orange, circle
     symbols) of MERP. We observe that $c^{\star}$ corresponding to
     version $V=1$ coincides with the critical average degree
     $\bar{c}$ for observing the MCGC on the same multiplex network.}
   \label{fig:c_star_M}
\end{figure}

This phenomenology can be observed in
Fig.~\ref{fig:phase_diagram}, where we plot the phase diagrams of
version $V=1$ and version $V=2$ of MERP for $M=2$, clearly
demonstrating the reentrant phase transition for $V=2$.
We can derive
the behavior of $p-p^{\star}$ versus $c-c^{\star}$ expanding the
Eqs. $H^{(2)}(S_1;p,c)=0$ around $p=p^{\star}$ and $c=c^{\star}$ with
$S_1=S_1^{\star}(p,c)$.
In this way, using Eq.\eqref{dis2} we obtain,
\bea
0=\frac{1}{2}\left.\frac{\partial^2 H^{(2)}}{\partial p^2}
\right|_{S_1^{\star},p^{\star},c^{\star}}
(\delta p)^2
+
\left. \frac{\partial H^{(2)}}{\partial c}
\right|_{S_1^{\star},p^{\star},c^{\star}}\delta c,
\eea
where $\delta p=p-p^{\star}$ and $\delta c=c-c^{\star}$.  Thus
if the signs of
$\left. {\partial ^2H^{(2)}}/{\partial p^2}\right|_{S_1^{\star},p^{\star},c^{\star}}$
and
$\left.{\partial H^{(2)}}/{\partial c}\right|_{S_1^{\star},p^{\star},c^{\star}}$
are opposite we obtain the scaling \bea |p-p^{\star}|\propto
(c-c^{\star})^{1/2}.  \eea Finally, in
Fig. \ref{fig:c_star_M} we show the different behavior of
$c^{\star}$ as a function of the number of layers $M$ for the version
$V=1$ as well as for the version $V=2$. We observe that as the number
of layer increases, the range of values of the average degree where the
reentrant phase is observed, increases significantly.

\section{Conclusions}
In conclusion, this work establishes a general theoretical framework for extended-range percolation in simple and multiplex networks, providing a relevant new class of exactly solvable percolation problems.
Extended-range
percolation on simple networks defines a percolation process in which communication between trusted nodes is ensured even if they are not directly connected. Specifically communication between two trusted nodes is allowed if  they are connected by paths involving exclusively trusted and untrusted facilitator nodes. These facilitator nodes, although untrusted, can still ensure communication between trusted nodes if they lie on at least a path of distance at most  $R$ between two trusted nodes. 
This percolation process reduces for $R=1$ to standard percolation and to standard interdependent percolation on single and multiplex networks, respectively.
Our theory builds on a message
passing approach providing the exact solution for the size of the extended-range giant
component (ERGC) for arbitrary finite $R$, as long as the network is locally tree-like.  Our message passing approach allows us to formulate a general theory for ERP on simple uncorrelated random networks. This general theory coincides with previous results obtained with a different approach for $R\leq 6$~\cite{cirigliano2023extended}. The resulting ERP transition is continuous, and characterized by a
percolation threshold that decreases as $R$ increases,
demonstrating the improved robustness of the network when this notion
of connectivity is adopted.  Furthermore, this general framework
allows us to {introduce and} study the multiplex extended-range percolation (MERP). This novel process enforces interdependencies
between trusted nodes, and {can be defined} in two variants {($V=1$ and $V=2$)} depending on {how interdependencies for untrusted nodes are considered}. We {provide the exact solution for the size of the} Multiplex Extended-Range Giant Component (MERGC) for arbitrary $R$. 
We show that the MERGC emerges with a discontinuous hybrid transition at a percolation threshold $p=p^{\star}$, which decreases as $R$ increase. Thus the phase diagram of MERP reflects a trade-off between the increased fragility {as a consequence of the multiplex network interdependencies}, and the increased robustness implied by the extended-range mechanism.
In particular we observe that version $V=2$, in which interdependencies are present only for trusted nodes, displays a rich phase diagram with the presence of a reentrant phase for some multiplex network topologies.

This work opens new perspectives to study the role of extended-range percolation in a variety of settings. At a fundamental level, it would
be interesting to explore the effect of design principles associating the state of trusted nodes to specific nodes of the network, for example, preferably to nodes of low or high degree. Additionally it would be very interesting to build on these results to formulate  more specific models which could be applied in realistic setting of quantum communications or of multilayer models of epidemic spreading.
More generally it would also be relevant to explore applications of this
framework to hypergraphs and  higher-order networks,
where several generalized percolation problems have been recently
proposed~\cite{bianconi2024theory,kim2024higher,sun2021higher,sun2023dynamic}.

\section*{Acknowledgments}
The authors thank G\'abor Tim\'ar for useful discussions
in the early stages of this project.

\appendix
\section{Message passing theory for extended-range percolation}
\label{ApMP}

The aim of this Appendix is to present a general
message-passing (MP) theory for extended-range percolation on single networks.
The MP formalism allows to express the equations  for extended-range percolation for arbitrary $R$ on single and  multiplex networks, as shown
in the main body of the paper.

In order to   formulate a message passing algorithm for extended-range percolation determining the giant component of the model,
we follow the general approach \cite{bianconi2018multilayer} for percolation problems. This consist by first deriving  the message passing equations when we know for each node $i$ if it is trusted ($x_i=1$) or untrusted ($x_i=0$), i.e. we know the exact configuration of trusted and untrusted nodes. Subsequently one can consider the scenario in which the exact configuration of trusted nodes is not known, and the only available information is that probability $p$ that a random node is trusted. \\
Let us now assume that we know the exact configuration of the trusted nodes, i.e. we have access to the variables $\{x_i\}$.
We consider two sets of messages
$\sigma_{i\to j}^r$ and $\omega_{i\to j}^r$, with $r \in \{1,2,\dots, R\}$,
associated to the directed links $i \to j$.

We consider two set of messages $\sigma_{i\to j}^r\in \{0,1\}$ and $\omega_{i\to j}^r\in \{0,1\}$,  with $r \in \{1,2,\dots, R\}$,
associated to the directed links $i \to j$.
Each message is a binary variable that can be $0$ or $1$, according to
the following rules.
The  messages  $\sigma_{i\to j}^{r+1}$ is one, i.e. $\sigma_{i\to j}^{r+1}=1$ when the node $i$ is connected to at least a trusted node in the ERGC by a path of length $r<R$ when the link $(i,j)$ is removed.
In all other scenarios we have $\sigma_{i\to j}^{r+1}=0$.\\
The messages $\omega_{i\to j}^{r+1}=1$ if is not connected to any trusted node in the ERGC but is connected to at least a trusted node  not in the ERGC by a path of length $r<R$ when the link $(i,j)$ is removed.
In all other scenarios we have $\omega_{i\to j}^{r+1}=0$.\\
Note that by construction at most one of the messages from $i$ to $j$ is different from zero; but it may also happen that all messages from $i$ to $j$ are zero. 

The equations for $\sigma_{i\to j}^1$ and $\omega_{i,j}^1$ are similar to the messages in standard percolation \cite{newman2023message,hartmann2006phase,bianconi2018multilayer}:
\bea
{\sigma}_{i\to j}^1&=&x_i\left[1-\prod_{\ell\in \partial i\setminus j}(1-\sum_{1\leq r\leq R}{\sigma}_{\ell\to i}^r)\right]\nonumber \\
{\omega}_{i\to j}^{1}&=&x_i\left[\prod_{\ell\in \partial i\setminus j}(1-\sum_{1\leq r\leq R}{\sigma}_{\ell\to i}^r)\right],
\eea
where here and in the following $\partial i$ denotes the neighbourhood of node $i$.

All other messages $\sigma_{i\to j}^r$ and $\omega_{i\to j}^r$ with $r>1$
are zero for links departing from trusted nodes.

\begin{figure}[!htb!]
   \includegraphics[width=0.48\textwidth]{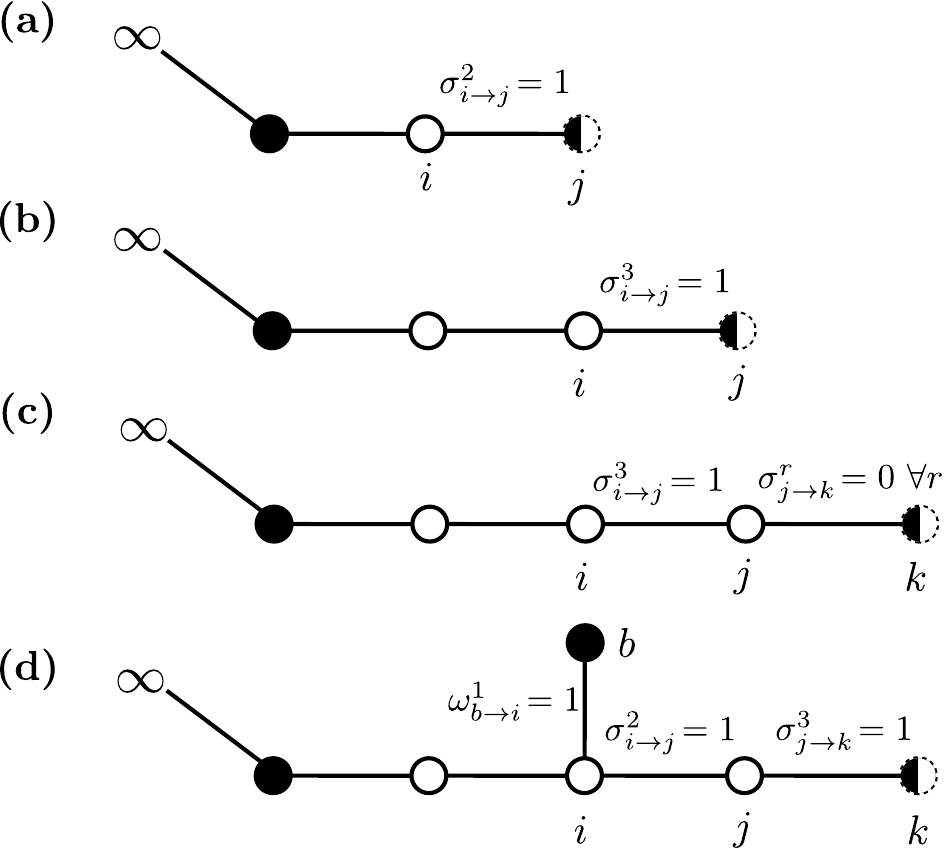}
   \caption{Schematic representation of the message passing algorithm for extended-range percolation with $R=3$, and a visualization of the role of bridge nodes. Trusted nodes are represented by filled circles, untrusted nodes by empty circles, and the dashed nodes can either be trusted or untrusted. Panels (a)-(c) describe the message passing algorithm in the {\em standard scenario}. Node $j$ in panel (a) is always part of the ERGC, while in (b) node $j$ belongs to the ERGC only if it is trusted. Node $k$ in panel (c) is not connected to the ERGC as it does not receive any positive message $\sigma_{j\to k}^r=0$ for $r \in \{ 1,2,3 \}$. Panel (d) describes the {\em bridge node scenario}. The presence of node $b$, sending the message $\omega_{b \to i }^1=1$, allows node $k$ -- if trusted -- to be part of the ERGC.}
   \label{fig:2}
\end{figure}

Concerning messages departing from untrusted nodes,
to write down recursive equations for $\sigma_{i\to j}^{r+1}$ with $r>0$.
it is  important to realize that two mutually exclusive scenarios may occur. We call these two scenarios the {\em standard scenario} and the {\em bridge node scenario}.
 In the {\em standard scenario}  node $i$ is connected to the ERGC through a node at distance $r$ from the closest trusted node and  node $i$ is at distance larger than $r+1$ from any other trusted trusted node.
 In the {\em bridge node scenario } node $i$ is connected to the ERGC through nodes that are a distance $r_1>r$ from trusted nodes in the ERGC. In this case the value of $r$ is determined by the presence of {\em bridge nodes}.
 A bridge node is a trusted node  that is not connected to the ERGC if node $i$ is removed, but it is connected to the ERGC if the node $i$ is not removed.  Thus assuming that the bridge node is  at distance $r$ for node $i$ we must impose $r+r_1\leq R$ as node $i$ need to act as a untrusted  facilitator node (for a schematic representation of the message passing algorithm and the role of bridge node see Fig. $\ref{fig:2}$).
These considerations can thus be used to determine the conditions under which the message $\sigma_{i\to j}^{r+1}=1$ is sent from node $i$ to node $j$.
Specifically we have that $\sigma_{i\to j}^{r+1}=1$ if node $i$ is untrusted and if either scenario (1) or scenario (2) occur (note that the two scenarios are mutually exclusive):
\begin{itemize}
\item[(1)]{\em Scenario bridge nodes}\\
\begin{itemize}
\item[(i)] node $i$ receives at least one message $\sigma_{\ell \to i}^{r_1}=1$ with $r_1>r$, and no messages $\sigma_{\ell \to i}^{r_2}=1$ with $r_2<r_1$
\item[(ii)] node $i$ receives at least one message $\omega_{\ell\to i}^r=1$ and no messages  $\omega_{\ell \to i}^{r_3}=1$ with $r_3<r$;
\item[(iii)] additionally we require $r+r_1\leq R$ or $r_1\leq R-r$. (This is to ensure that the bridge node is a {\em facilitator}, i.e.  connected to the ERGC, or equivalently that node $i$ is in a active path between trusted nodes). Requiring the existence of at least a given $r_1$ with  $r<r_1\leq R-r$ we need to impose $R-r>r$ or equivalently $R-2r>0$ for allowing scenario (1) to hold.
\end{itemize}
\item[(2)] {\em Standard scenario:}\\
\begin{itemize}
\item[(a)]
node $i$ receives at least one message $\sigma_{\ell \to i}^r=1$ with $r<R$ and it does not receive any message $\sigma_{\ell \to i}^{r_1}=1$  with $r_1<r$;
\item [(b)]
node $i$ does not receive any message $\omega_{\ell \to i}^{r_2}=1$ that has $r_2\leq \bar{r}=\min(r-1,R-r)$ . Indeed we need to exclude the messages $\omega_{\ell \to i}^{r_2}=1$ with $r_2<r$ and $r_2\leq R-r$ that could act as bridge nodes and change the $r$ of the message as in the bridge node scenario.
\end{itemize}
\end{itemize}

This leads to the following equations:
\begin{widetext}
\bea
&&{\sigma}_{i\to j}^{r+1}=(1-x_i) \left\{\left[\prod_{\ell\in \partial i\setminus j}(1-\sum_{1\leq q\leq{r-1} }{\sigma}_{\ell\to i}^q-\sum_{1\leq q\leq\bar{r}}{\omega}_{\ell\to i}^q)-\prod_{\ell\in \partial i\setminus j}(1-\sum_{1\leq q\leq r}{\sigma}_{\ell\to i}^q-\sum_{1\leq q\leq \bar{r}}{\omega}_{\ell\to i}^q)\right]\right.\nonumber \\
&&+\theta({R-2r})\left[\prod_{\ell\in \partial i\setminus j}(1-\sum_{1\leq q\leq r}{\sigma}_{\ell\to i}^q-\sum_{1\leq q\leq r-1}{\omega}_{\ell\to i}^q)+\prod_{\ell\in \partial i\setminus j}(1-\sum_{1\leq q\leq   R-r}{\sigma}_{\ell\to i}^q-\sum_{1\leq q\leq r}{\omega}_{\ell\to i}^q)\right.\nonumber \\
&&\left.\left.-\prod_{\ell\in \partial i\setminus j}(1-\sum_{1\leq q<  r}{\sigma}_{\ell\to i}^q-\sum_{1\leq q\leq r}{\omega}_{\ell\to i}^q)
-\prod_{\ell\in \partial i\setminus j}(1-\sum_{1\leq q\leq  R-r}{\sigma}_{\ell\to i}^q-\sum_{1\leq q\leq r-1}{\omega}_{\ell\to i}^q)\right]\right\}.
\label{mess_sigma}
\eea
\end{widetext}
Note that here and in the following we use $\theta(x)$ to indicate the Heaviside function with $\theta(x)=1$ if $x>0$ and $\theta(x)=0$ otherwise.
The first row describes scenario (2), the rest of the equations describes scenario (1) and the expression is a direct consequence of the inclusion-exclusion principle.\\
In particular the expression describing scenario (1) can be deduced by considering the expression valid as long as $\theta({R-2r})=1$
\bea
\left[\prod_{\ell\in \partial i\setminus j}(1-\sum_{1\leq q\leq r-1}{\omega}_{\ell\to i}^q)-\prod_{\ell\in \partial i\setminus j}(1-\sum_{1\leq q\leq r}{\omega}_{\ell\to i}^q)\right]\nonumber \\
\times \left[\prod_{\ell\in \partial i\setminus j}(1-\sum_{1\leq q\leq r}{\sigma}_{\ell\to i}^q)-\prod_{\ell\in \partial i\setminus j}(1-\sum_{1\leq q\leq  R-r}{\sigma}_{\ell\to i}^q)\right].\nonumber
\eea
Here we assume that  $\theta({R-2r})=1$ implementing in part condition (iii), while the two multiplicative terms  indicate the condition (ii) and the conditions [(i),(iii)] respectively.
Since any product of $\omega_{\ell\to i}^r\sigma_{\ell\to i}^{r_1}=0$ for all $r$ and $r_1$ (due to the fact that the messages are non-zero in mutually exclusive situations) we recover that the above expression is equivalent to  the inclusion-exclusion principle expression:
\bea
&&\left.\prod_{\ell\in \partial i\setminus j}(1-\sum_{1\leq q\leq r}{\sigma}_{\ell\to i}^q-\sum_{1\leq q\leq r-1}{\omega}_{\ell\to i}^q)\right.\nonumber\\ &&+\prod_{\ell\in \partial i\setminus j}(1-\sum_{1\leq q\leq  R-r}{\sigma}_{\ell\to i}^q-\sum_{1\leq q\leq r}{\omega}_{\ell\to i}^q)\nonumber \\
&&-\prod_{\ell\in \partial i\setminus j}(1-\sum_{1\leq q<  r}{\sigma}_{\ell\to i}^q-\sum_{1\leq q\leq r}{\omega}_{\ell\to i}^q)\nonumber \\
&&\left.-\prod_{\ell\in \partial i\setminus j}(1-\sum_{1\leq q\leq  R-r}{\sigma}_{\ell\to i}^q-\sum_{1\leq q\leq r-1}{\omega}_{\ell\to i}^q)\right..\nonumber
\eea
This is exactly the expression that multiplies the factor $\theta(R-2r)$ in the message passing equation for $\sigma_{i\to j}^{r+1}$ and that implements the bridge node scenario.
\\
Having derived the equation for the message $\sigma_{i\to j}^{r+1}$ we now derive the equation for the message  $\omega_{i\to j}^{r+1}$.
The untrusted facilitator node $i$ sends a message $\omega_{i\to j}^{r+1}=1$ with $r>0$ if:
\begin{itemize}
\item
it does not  receive any positive message $\sigma_{\ell\to i}^{r_1}$ with $1\leq r_1<R$;
\item
it receives at least one positive message $\omega_{\ell \to i}^r=1$ and no messages $\omega_{\ell\to i}^{r_1}=1$ with $r_1<r$.
\end{itemize}
This leads then to the message passing equations:
\bea
{\omega}_{i\to j}^{r+1}&=&(1-x_i) \left[\prod_{\ell\in \partial i\setminus j}(1-\sum_{1\leq r \leq R-1}{\sigma}_{\ell\to i}^r-\sum_{1\leq q\leq r-1}\hat{\omega}_{\ell\to i}^q)\right.\nonumber \\
&&\left.-\prod_{\ell\in \partial i\setminus j}(1-\sum_{1\leq r \leq R-1}{\sigma}_{\ell\to i}^r-\sum_{1\leq q\leq r}\hat{\omega}_{\ell\to i}^q)\right].
\label{mess_w}
\eea
The fraction $P^{\infty}$ of  nodes that are trusted and are in the ERGC is then given by
\bea
P^{\infty}=\frac{1}{N}\sum_{i=1}^N x_i\left[1-\prod_{\ell\in \partial i}(1-\sum_{1\leq r\leq R}{\sigma}_{\ell\to i}^r)\right].
\label{Pinf_xi}
\eea
while the fraction $U^{\infty}$ of nodes that are untrusted and are in the ERGC is given by
\bea
U^{\infty}=\frac{1}{N}\sum_{i=1}^N (1-x_i)\left[1-\prod_{\ell\in \partial i}(1-\sum_{1\leq r\leq R-1}{\sigma}_{\ell\to i}^r)\right].
\label{Uinf_xi}
\eea
The message passing equations \eqref{mess_sigma} and \eqref{mess_w} fully determine the order parameter $P^{\infty}$ given by Eq. \eqref{Pinf_xi} when we know the configuration of trusted and untrusted nodes, i.e. the configuration $\{x_i\}$.
However in a number of cases the exact configuration $\{x_i\}$ determining which nodes are trusted and which are untrusted is not know. In this scenario we can assume that the state of a node $x_i$ is drawn independently at random with a probability $p$ that the node is trusted, thus we assume that the probability $\tilde{P}(\{x_i\})$ of a configuration is given by \bea
\tilde{P}(\{x_i\})=\prod_{i=1}^Np^{x_i}(1-p)^{1-x_i}.
\eea   In this case we need to modify the message passing equations by averaging over the probability $\tilde{P}(\{x_i\})$. This alternative message passing algorithm is formulated in terms of a new set of   messages: the messages $\hat{\sigma}_{i\to j}^r\in [0,1]$ indicating  the probability that a  node $i$  is connected to the ERGC by nodes different from $j$ and at distance $r$ from active trusted  nodes and the messages $\hat{\omega}_{i\to j}^r\in [0,1]$ indicating  the probability that a  node $i$ is not connected to the ERGC by nodes different from $j$ and is at distance $r$ from other trusted nodes. We have that $\hat{\sigma}_{i\to j}^r$ is the average of $\sigma_{i\to j}^r$ and $\hat{\omega}_{i\to j}^r$ is the average of $\omega_{i\to j}^r$ over the distribution $\tilde{P}(\{{\bf x}\})$ of trusted and untrusted nodes.\\
The messages $\hat{\sigma}_{i\to j}^r$ for $1\leq r\leq R$ obey
\begin{widetext}
\bea
\hat{\sigma}_{i\to j}^1&=&p\left[1-\prod_{\ell\in \partial i\setminus j}(1-\sum_{1\leq r\leq R}\hat{\sigma}_{\ell\to i}^r)\right]\nonumber \\
\hat{\sigma}_{i\to j}^{r+1}&=&(1-p) \left\{\left[\prod_{\ell\in \partial i\setminus j}(1-\sum_{1\leq q\leq r-1}\hat{\sigma}_{\ell\to i}^q-\sum_{1\leq q\leq\bar{r}}\hat{\omega}_{\ell\to i}^q)-\prod_{\ell\in \partial i\setminus j}(1-\sum_{1\leq q\leq r}\hat{\sigma}_{\ell\to i}^q-\sum_{1\leq q\leq\bar{r}}\hat{\omega}_{\ell\to i}^q)\right]\right.\nonumber \\
&&+\theta({R-2r})\left[\prod_{\ell\in \partial i\setminus j}(1-\sum_{1\leq q\leq r}\hat{\sigma}_{\ell\to i}^q-\sum_{1\leq q\leq r-1}\hat{\omega}_{\ell\to i}^q)+\prod_{\ell\in \partial i\setminus j}(1-\sum_{1\leq q\leq  R-r}\hat{\sigma}_{\ell\to i}^q-\sum_{1\leq q\leq r}\hat{\omega}_{\ell\to i}^q)\right.\nonumber \\
&&\left.\left.-\prod_{\ell\in \partial i\setminus j}(1-\sum_{1\leq q<  r}\hat{\sigma}_{\ell\to i}^q-\sum_{1\leq q\leq r}\hat{\omega}_{\ell\to i}^q)
-\prod_{\ell\in \partial i\setminus j}(1-\sum_{1\leq q\leq  R-r}\hat{\sigma}_{\ell\to i}^q-\sum_{1\leq q\leq r-1}\hat{\omega}_{\ell\to i}^q)\right]\right\}.
\eea
\end{widetext}
The messages $\hat{\omega}_{i\to j}^r$  with $1 \leq r \leq \ceil{R/2}-1$ obey
\begin{widetext}
\bea
\hat{\omega}_{i \to j}^{1} &=& p \left[\prod_{\ell \in \partial i \setminus j} (1-\sum_{1 \leq q \leq R}\hat{\sigma}_{\ell \to i}^r)\right],\nonumber \\
\hat{\omega}_{i \to j}^{r+1} &=& (1-p) \left[\prod_{\ell \in \partial i \setminus j}(1-\sum_{1 \leq q \leq R-1}\hat{\sigma}_{\ell \to i}^q-\sum_{1 \leq q \leq r-1}\hat{\omega}_{\ell \to i}^q)-\prod_{\ell \in \partial i \setminus j}(1-\sum_{1 \leq q \leq R-1}\hat{\sigma}_{\ell \to i}^r-\sum_{1\leq q\leq r}\hat{\omega}_{\ell \to i}^q)\right].\nonumber\\
\eea
\end{widetext}

Note that in total the messages are $R+\ceil{R/2}-1$. The order parameter $P^{\infty}$ expressing the fraction of nodes that are trusted and are in the  ERGC is given by
\bea
P^{\infty}=\frac{p}{N}\sum_{i=1}^N\left[1-\prod_{\ell\in \partial i}(1-\sum_{1\leq r\leq R}\hat{\sigma}_{\ell\to i}^r)\right],
\eea
while the fraction of nodes $U^{\infty}$ that are untrusted and are in the ERGC is given by
\bea
U^{\infty}=\frac{(1-p)}{N}\sum_{i=1}^N\left[1-\prod_{\ell\in \partial i}(1-\sum_{1\leq r\leq R-1}\hat{\sigma}_{\ell\to i}^r)\right].
\eea

These equations are the starting point to formulate the equation determining the size of the ERGC on random networks with given degree distribution. To this end we consider networks $\mathcal{G}=(V,E)$  drawn from the distribution
\bea
\mathcal{P}(\mathcal{G})=\prod_{i=1}^{N}\prod_{j=i}^N p_{ij}^{a_{ij}}(1-p_{ij})^{1-a_{ij}}
\eea
where ${\bf a}$ indicates the adjacency matrix of the network and  $p_{ij}$ indicates the probability of link between nodes $i$ and $j$. The probability $p_{ij}$ is expressed in terms of the degrees $k_i$ and $k_j$ associated respectively to the two nodes $i$ and $j$, as
\bea
p_{ij}=\frac{k_ik_j}{\Avg{k}N}.
\eea
Here $P(k)$ indicates the degree distribution, i.e. fraction of nodes of degree $k$.
In order to derive the general equations determining the size of the ERGC,  we average the message passing equations over the distribution $\mathcal{P}(\mathcal{G})$.
In this case we indicate with $S_r$ the average of  $\hat{\sigma}_{i\to j}^r$ and  $W_r$ is the average of $\hat{\omega}_{i\to j}^r$ over the probability $\mathcal{P}(\mathcal{G})$. In this way we derive the equations Eq.\eqref{mes_ensemble}.
Moreover we indicate with $P^{\infty}$ (and with $U^{\infty}$) the probability that  node is trusted (untrusted) and  belongs to the ERGC obtaining Eqs.\eqref{Pinf_c} and \eqref{Uinf_c}.

\section{Equations for ERP for finite $R$}
\label{AppendixA}
In this Appendix we provide the explicit equations  \eqref{mes_ensemble} for extended-range percolation with $R\in\{1,2,3,4\}$  on a simple network. 
\subsection{The $R=2$ equations}
\label{R2}
For $R=2$ the self-consistent equations involve two average messages, $S_1$ and $S_2$ and read
\bea
S_1&=&p [1-G_1(1-S_1-S_2)], \nonumber \\
S_2&=&(1-p) [1-G_1(1-S_1)].
\label{eq:R2_m}
\eea
The order parameter $P^{\infty}$ is given by
\bea
P^{\infty}=p[1-G_0(1-S_1-S_2)],
\label{eq:P2}
\eea
while $U^{\infty}$ if given by 
\bea
U^{\infty}=(1-p)[1-G_0(1-S_1)].
\eea
\subsection{The $R=3$ equations}
\label{R3}
For $R=3$ the self-consistent equations involve four average messages, $S_1,S_2,S_3$ and $W_1$ reflecting that for $R>2$ the effect to bridge nodes needs to be taken into account. These equations  read
\begin{widetext}
\bea
S_1&=&p [1-G_1(1-S_1-S_2-S_3)], \nonumber \\
S_2&=&(1-p) [1 + G_1(1-S_1-S_2-W_1)-G_1(1-S_1-W_1)-G_1(1-S_1-S_2)],\nonumber \\
S_3&=&(1-p) [G_1(1-S_1-W_1)-G_1(1-S_1-S_2-W_1)],\nonumber \\
W_1&=&pG_1(1-S_1-S_2-S_3).
\label{eq:R3_m}
\eea
The order parameter $P^{\infty}$ is given by
\bea
P^{\infty}=p[1-G_0(1-S_1-S_2-S_3)],
\label{eq:P3}
\eea
while $U^{\infty}$ if given by 
\bea
U^{\infty}=(1-p)[1-G_0(1-S_1-S_2)].
\eea
\subsection{The $R=4$ equations }
\label{R4}
For $R=4$ the self-consistent equations involve five average messages, $S_1,S_2,S_3,S_4$ and $W_1$ and read
\bea
S_1&=&p [1- G_1 (1-S_1-S_2-S_3-S_4)], \nonumber \\
S_2&=&(1-p) [1 + G_1 (1-S_1-S_2-S_3-W_1)-G_1(1-S_1-W_1)-G_1(1-S_1-S_2-S_3)],\nonumber \\
S_3&=&(1-p) [G_1(1-S_1-W_1)-G_1(1-S_1-S_2-W_1)],\nonumber \\
S_4&=&(1-p) [G_1(1-S_1-S_2-W_1)-G_1(1-S_1-S_2-S_3-W_1)],\nonumber \\
W_1&=&p G_1 (1-S_1-S_2-S_3-S_4). \label{eq:R4_m}
\eea
The order parameter $P^{\infty}$ is given by
\bea
P^{\infty}&=&p [1-G_0 (1-S_1-S_2-S_3-S_4)],
\label{eq:P4}
\eea
while $U^{\infty}$ if given by 
\bea
U^{\infty}=(1-p)[1-G_0(1-S_1-S_2-S_3)].
\eea

\end{widetext}

\section{Equivalence of Eqs. \eqref{mes_ensemble} with the equations derived in Ref. \cite{cirigliano2023extended} for $R\leq 4$.}
\label{AppendixERP}

In this Appendix we demonstrate the equivalence of the equations \eqref{mes_ensemble} for extended-range percolation with $R \leq 4$ discussed in Appendix \ref{AppendixA} with the equations previously derived in Ref. \cite{cirigliano2023extended} using a different formalism. Note that in Ref. \cite{cirigliano2023extended} the equations were derived explicitly up to $R=6$. All these equations are equivalent to the equations derived in this work. However, the expressions are rather combinatorially involved and we prefer for the sake of simplicty to prove the equivalence between  only with $R \leq 4$.

According to the approach of \cite{cirigliano2023extended}, the  equations for ERP on random networks with $R = 4$ are
\bea
u_1 &=& G_1\big(p u_1 +(1-p)u_2), \nonumber \\
u_2 &=& G_1\big(p u_1 +(1-p)u_2), \nonumber \\
u_3 &=& G_1\big((1-p)u_4\big) - G_1\big((1-p)u_3\big)+u_2, \nonumber \\
u_4 &=& G_1(1-p) -G_1\big((1-p)u_3\big)+u_2,\nonumber \\
P^{\infty}&=&p\left[1-G_0\big(pu_1+(1-p)u_2)\right], \label{eq:ERP_first_paper}
\eea
where $u_1,\dots,u_4$ are probabilities of not reaching the ERGC following a randomly chosen link conditioned to different configurations, e.g. for $u_1$ the chosen link ends in a trusted node, for $u_2$ the chosen link emanates from a trusted node and ends in an untrusted node, while $u_3$ and $u_4$ take care of more complex combinatoric  configurations including bridge nodes.
For the interested reader, we defer to \cite{cirigliano2023extended} for the details of such formalism. For the purpose of this Appendix, it is sufficient to know that the $\{u_r\}$ are probabilities and that they have a different physical interpretation from the $\{S_r,W_r\}$. The equations for $R=3$ are recovered by simply setting $u_4=1$. Analogously, setting $u_3=1$ we recover the equations for $R=2$, and setting $u_2=1$ finally gives the equation for standard site percolation ($R=1$).
Note that the order parameter $U^{\infty}$ had not been introduced in \cite{cirigliano2023extended}.

\subsection{The $R=2$ equations}
Let us now prove the equivalence between the Eqs. (\ref{eq:R2_m})-(\ref{eq:P2}) and the equations (\ref{eq:ERP_first_paper}) with $u_3=u_4=0$.
To define a mapping between the two sets of variables $S_r$ and $u_r$,  we proceed as follows.
First, comparing the two different equations for $P^{\infty}$ written in the two sets of variables, we get the condition
\[1-S_1-S_2 = pu_1 +(1-p)u_2. \]
Then, from the equation for $u_1$ and a comparison with the equation for $S_1$, we get
\[S_1=p(1-u_1). \]
The solution of these equations is
\bea
S_1&=&p(1-u_1),\nonumber \\
S_2&=&(1-p)(1-u_2),
\eea
and shows the equivalence of the two formalisms.
\subsection{The $R=3$ equations}
Let us now prove the equivalence between the Eqs. (\ref{eq:R3_m})-(\ref{eq:P3}) and the equations (\ref{eq:ERP_first_paper}) with $u_4=0$.
Reasoning as we did for $R=2$ in the previous subsection, from the two equations for $P^{\infty}$ we get the condition
\bea
1-S_1-S_2-S_3 = pu_1 +(1-p)u_2.\label{eq:condition_R_3}
\eea
Using then the equations for $u_1$ and $S_1$, we get again $S_1=p(1-u_1).$
The equation for $W_1$ gives instead $W_1=p u_1$. From \eqref{eq:condition_R_3} we get
\[1-p-S_2-S_3 = (1-p)u_2. \]
Using then the two equations for $S_2$ and $S_3$, after some simplifications this equation can be rewritten as
\[u_2=G_1\big(1-S_1-S_2 \big). \]
Comparing this equation with the equation for $u_2$ in \eqref{eq:ERP_first_paper} we finally get
\[G_1(pu_1+1-p-S_2) = G_1\big(pu_1 +(1-p)u_3 \big), \]
from which $S_2=(1-p)(1-u_3)$. The equivalence between the two formalisms for $R=3$ is then proved via the mapping
\bea
S_1&=&p(1-u_1),\nonumber \\
S_2&=&(1-p)(1-u_3), \nonumber \\
S_3&=& (1-p)(u_3-u_2), \nonumber \\
W_1&=&pu_1.
\eea

\subsection{The $R=4$ equations}
Let us now prove the equivalence between the Eqs. (\ref{eq:R4_m})-(\ref{eq:P3}) and the equations (\ref{eq:ERP_first_paper}). 
In perfect analogy to what we did in the two previous subsections, we can determine a mapping between the $\{S_r,W_r\}$ and the $\{u_r\}$ variables to show the equivalence of the two sets of equations.
The first step is to compare the equations for $P^{\infty}$, to get the condition
\bea
1-S_1-S_2-S_3-S_4 = pu_1 +(1-p)u_2,\label{eq:condition_R_4}
\eea
which, together with the equations for $S_1$ and $W_1$ determines again $S_1=p(1-u_1)=p-W_1$.
Eq.~\eqref{eq:condition_R_4} then gives
\[1-p-S_2-S_3-S_4 = (1-p)u_2, \]
which can be rewritten after some simplifications, using the equations for $S_2,S_3,S_4$, as
\[u_2=G_1(pu_1 + 1-p -S_2-S_3). \]
By comparison with the equation for $u_2$ we get
\[1-p-S_2-S_3=(1-p)u_3, \]
which can be in turn rewritten after some simplifications, using again the equations for $S_2$,$S_3$ and the expressions derived above, as
\[u_3=\left[G_1(1-p-S_2)-G_1\big((1-p)u_3\big)+u_2 \right]. \]
By comparison with the equation for $u_3$ we finally get $S_2=(1-p)(1-u_4)$. Hence we proved the equivalence of the two formalisms for $R=4$ via the mapping
\bea
S_1&=&p(1-u_1),\nonumber \\
S_2&=&(1-p)(1-u_4), \nonumber \\
S_3&=& (1-p)(u_3-u_2), \nonumber \\
S_4&=&(1-p)(u_4-u_3),\nonumber \\
W_1&=&pu_1.
\eea
\section{Equations for MERP for finite $R\leq 4$}
\label{AppendixMERP}

In this Appendix we write down explicitly the equations for multiplex extended-range percolation \eqref{mes},~\eqref{eq:P_inf_MERP},~\eqref{eq:U_inf_MERP1}, and \eqref{eq:U_inf_MERP2} for $R \leq 4$ on multiplex uncorrelated random graphs. We assume that the degrees of the same node across different layers are uncorrelated. Under such hypothesis, we have that
$S_r^{\alpha}=S_r,\quad W_r^{\alpha}=W_r$, thus the equations greatly simplify.

\subsection{The $R=1$ equations}
The equations in this case involve only the message $S_1$, and they reduce for both versions -- since untrusted nodes are not considered at all -- to the standard MCGC equations for MIP, \cite{bianconi2018multilayer}. In this case, the order parameter $U^{\infty}$ is identically zero.

\subsection{The $R=2$ equations}
For $R=2$ the self-consistent equations involve two average messages, $S_1$ and $S_2$ and read
\bea
S_1&=&p [1-G_1(1-S_1-S_2)][1-G_0(1-S_1-S_2)]^{M-1}, \nonumber \\
S_2&=&(1-p) [1-G_1(1-S_1)]\nonumber \\
&&\times\{\delta_{V,2}+\delta_{V,1}[1-G_0(1-S_1)]^{M-1}\}.
\eea
The order parameter $P^{\infty}$ is given by
\bea
P^{\infty}=p[1-G_0(1-S_1-S_2)]^M,
\eea
while $U^{\infty}$ is given by
\bea
U^{\infty}&=&(1-p)[1-G_0(1-S_1)]\nonumber \\
&&\times\{\delta_{V,2}+\delta_{V,1}[1-G_0(1-S_1)]^{M-1}\}.
\eea
\subsection{The $R=3$ equations}
For $R=3$ the self-consistent equations involve four average messages, $S_1,S_2,S_3$ and $W_1$ reflecting that for $R>2$ the effect to bridge nodes needs to be taken into account. These equations  read
\begin{widetext}
\bea
S_1&=&p [1-G_1(1-S_1-S_2-S_3)][1-G_0(1-S_1-S_2-S_3)]^{M-1}, \nonumber \\
S_2&=&(1-p) [1 + G_1(1-S_1-S_2-W_1)-G_1(1-S_1-W_1)-G_1(1-S_1-S_2)]\{\delta_{V,2}+\delta_{V,1}[1-G_0(1-S_1-S_2)]^{M-1}\},\nonumber \\
S_3&=&(1-p) [G_1(1-S_1-W_1)-G_1(1-S_1-S_2-W_1)]\{\delta_{V,2}+\delta_{V,1}[1-G_0(1-S_1-S_2)]^{M-1}\},\nonumber \\
W_1&=&pG_1(1-S_1-S_2-S_3)[1-G_0(1-S_1-S_2-S_3)]^{M-1}.
\eea
The order parameter $P^{\infty}$ is given by
\bea
P^{\infty}=p[1-G_0(1-S_1-S_2-S_3)]^M,
\eea
while $U^{\infty}$ is given by
\bea
U^{\infty}=(1-p)[1-G_0(1-S_1-S_2)]\{\delta_{V,2}+\delta_{V,1}[1-G_0(1-S_1-S_2)]^{M-1}\}.
\eea

\subsection{The $R=4$ equations}
For $R=4$ the self-consistent equations involve five average messages, $S_1,S_2,S_3,S_4$ and $W_1$ and read

\bea
S_1&=&p [1- G_1 (1-S_1-S_2-S_3-S_4)][1- G_0 (1-S_1-S_2-S_3-S_4)]^{M-1}, \nonumber \\
S_2&=&(1-p) [1 + G_1 (1-S_1-S_2-S_3-W_1)-G_1(1-S_1-W_1)-G_1(1-S_1-S_2-S_3)]\nonumber \\
&&\times\{\delta_{V,2}+\delta_{V,1}[1-G_0(1-S_1-S_2+S3)]^{M-1}\},\nonumber \\
S_3&=&(1-p) [G_1(1-S_1-W_1)-G_1(1-S_1-S_2-W_1)]\{\delta_{V,2}+\delta_{V,1}[1-G_0(1-S_1-S_2+S3)]^{M-1}\},\nonumber \\
S_4&=&(1-p) [G_1(1-S_1-S_2-W_1)-G_1(1-S_1-S_2-S_3-W_1)]\{\delta_{V,2}+\delta_{V,1}[1-G_0(1-S_1-S_2+S3)]^{M-1}\},\nonumber \\
W_1&=&p G_1 (1-S_1-S_2-S_3-S_4)[1- G_0 (1-S_1-S_2-S_3-S_4)]^{M-1}. \eea
The order parameter $P^{\infty}$ is given by
\bea
P^{\infty}&=&p [1-G_0 (1-S_1-S_2-S_3-S_4)][1- G_0 (1-S_1-S_2-S_3-S_4)]^{M-1},
\eea
while $U^{\infty}$ is given by
\bea
U^{\infty}=(1-p)[1-G_0(1-S_1-S_2-S_3)]\{\delta_{V,2}+\delta_{V,1}[1-G_0(1-S_1-S_2+S_3)]^{M-1}\}.
\eea

\end{widetext}

\bibliography{references.bib}

\begin{thebibliography}{51}%
\makeatletter
\providecommand \@ifxundefined [1]{%
 \@ifx{#1\undefined}
}%
\providecommand \@ifnum [1]{%
 \ifnum #1\expandafter \@firstoftwo
 \else \expandafter \@secondoftwo
 \fi
}%
\providecommand \@ifx [1]{%
 \ifx #1\expandafter \@firstoftwo
 \else \expandafter \@secondoftwo
 \fi
}%
\providecommand \natexlab [1]{#1}%
\providecommand \enquote  [1]{``#1''}%
\providecommand \bibnamefont  [1]{#1}%
\providecommand \bibfnamefont [1]{#1}%
\providecommand \citenamefont [1]{#1}%
\providecommand \href@noop [0]{\@secondoftwo}%
\providecommand \href [0]{\begingroup \@sanitize@url \@href}%
\providecommand \@href[1]{\@@startlink{#1}\@@href}%
\providecommand \@@href[1]{\endgroup#1\@@endlink}%
\providecommand \@sanitize@url [0]{\catcode `\\12\catcode `\$12\catcode
  `\&12\catcode `\#12\catcode `\^12\catcode `\_12\catcode `\%12\relax}%
\providecommand \@@startlink[1]{}%
\providecommand \@@endlink[0]{}%
\providecommand \url  [0]{\begingroup\@sanitize@url \@url }%
\providecommand \@url [1]{\endgroup\@href {#1}{\urlprefix }}%
\providecommand \urlprefix  [0]{URL }%
\providecommand \Eprint [0]{\href }%
\providecommand \doibase [0]{https://doi.org/}%
\providecommand \selectlanguage [0]{\@gobble}%
\providecommand \bibinfo  [0]{\@secondoftwo}%
\providecommand \bibfield  [0]{\@secondoftwo}%
\providecommand \translation [1]{[#1]}%
\providecommand \BibitemOpen [0]{}%
\providecommand \bibitemStop [0]{}%
\providecommand \bibitemNoStop [0]{.\EOS\space}%
\providecommand \EOS [0]{\spacefactor3000\relax}%
\providecommand \BibitemShut  [1]{\csname bibitem#1\endcsname}%
\let\auto@bib@innerbib\@empty
\bibitem [{\citenamefont {Dorogovtsev}\ \emph {et~al.}(2008)\citenamefont
  {Dorogovtsev}, \citenamefont {Goltsev},\ and\ \citenamefont
  {Mendes}}]{dorogovtsev2008critical}%
  \BibitemOpen
  \bibfield  {author} {\bibinfo {author} {\bibfnamefont {S.~N.}\ \bibnamefont
  {Dorogovtsev}}, \bibinfo {author} {\bibfnamefont {A.~V.}\ \bibnamefont
  {Goltsev}},\ and\ \bibinfo {author} {\bibfnamefont {J.~F.~F.}\ \bibnamefont
  {Mendes}},\ }\bibfield  {title} {\bibinfo {title} {Critical phenomena in
  complex networks},\ }\href {https://doi.org/10.1103/RevModPhys.80.1275}
  {\bibfield  {journal} {\bibinfo  {journal} {Rev. Mod. Phys.}\ }\textbf
  {\bibinfo {volume} {80}},\ \bibinfo {pages} {1275} (\bibinfo {year}
  {2008})}\BibitemShut {NoStop}%
\bibitem [{\citenamefont {Li}\ \emph {et~al.}(2021)\citenamefont {Li},
  \citenamefont {Liu}, \citenamefont {Lü}, \citenamefont {Hu}, \citenamefont
  {Xu},\ and\ \citenamefont {Zhang}}]{li2021percolation}%
  \BibitemOpen
  \bibfield  {author} {\bibinfo {author} {\bibfnamefont {M.}~\bibnamefont
  {Li}}, \bibinfo {author} {\bibfnamefont {R.-R.}\ \bibnamefont {Liu}},
  \bibinfo {author} {\bibfnamefont {L.}~\bibnamefont {Lü}}, \bibinfo {author}
  {\bibfnamefont {M.-B.}\ \bibnamefont {Hu}}, \bibinfo {author} {\bibfnamefont
  {S.}~\bibnamefont {Xu}},\ and\ \bibinfo {author} {\bibfnamefont {Y.-C.}\
  \bibnamefont {Zhang}},\ }\bibfield  {title} {\bibinfo {title} {Percolation on
  complex networks: Theory and application},\ }\href
  {https://doi.org/https://doi.org/10.1016/j.physrep.2020.12.003} {\bibfield
  {journal} {\bibinfo  {journal} {Physics Reports}\ }\textbf {\bibinfo {volume}
  {907}},\ \bibinfo {pages} {1} (\bibinfo {year} {2021})}\BibitemShut {NoStop}%
\bibitem [{\citenamefont {Lee}\ \emph {et~al.}(2018)\citenamefont {Lee},
  \citenamefont {Kahng}, \citenamefont {Cho}, \citenamefont {Goh},\ and\
  \citenamefont {Lee}}]{lee2018recent}%
  \BibitemOpen
  \bibfield  {author} {\bibinfo {author} {\bibfnamefont {D.}~\bibnamefont
  {Lee}}, \bibinfo {author} {\bibfnamefont {B.}~\bibnamefont {Kahng}}, \bibinfo
  {author} {\bibfnamefont {Y.}~\bibnamefont {Cho}}, \bibinfo {author}
  {\bibfnamefont {K.-I.}\ \bibnamefont {Goh}},\ and\ \bibinfo {author}
  {\bibfnamefont {D.-S.}\ \bibnamefont {Lee}},\ }\bibfield  {title} {\bibinfo
  {title} {Recent advances of percolation theory in complex networks},\ }\href
  {https://doi.org/https://doi.org/10.3938/jkps.73.152} {\bibfield  {journal}
  {\bibinfo  {journal} {Journal of the Korean Physical Society}\ }\textbf
  {\bibinfo {volume} {73}},\ \bibinfo {pages} {152} (\bibinfo {year}
  {2018})}\BibitemShut {NoStop}%
\bibitem [{\citenamefont {Meng}\ \emph {et~al.}(2023)\citenamefont {Meng},
  \citenamefont {Hu}, \citenamefont {Tian}, \citenamefont {Dong}, \citenamefont
  {Lambiotte}, \citenamefont {Gao},\ and\ \citenamefont
  {Havlin}}]{meng2023percolation}%
  \BibitemOpen
  \bibfield  {author} {\bibinfo {author} {\bibfnamefont {X.}~\bibnamefont
  {Meng}}, \bibinfo {author} {\bibfnamefont {X.}~\bibnamefont {Hu}}, \bibinfo
  {author} {\bibfnamefont {Y.}~\bibnamefont {Tian}}, \bibinfo {author}
  {\bibfnamefont {G.}~\bibnamefont {Dong}}, \bibinfo {author} {\bibfnamefont
  {R.}~\bibnamefont {Lambiotte}}, \bibinfo {author} {\bibfnamefont
  {J.}~\bibnamefont {Gao}},\ and\ \bibinfo {author} {\bibfnamefont
  {S.}~\bibnamefont {Havlin}},\ }\bibfield  {title} {\bibinfo {title}
  {Percolation theories for quantum networks},\ }\href
  {https://doi.org/https://doi.org/10.3390/e25111564} {\bibfield  {journal}
  {\bibinfo  {journal} {Entropy}\ }\textbf {\bibinfo {volume} {25}},\ \bibinfo
  {pages} {1564} (\bibinfo {year} {2023})}\BibitemShut {NoStop}%
\bibitem [{\citenamefont {Artime}\ \emph {et~al.}(2024)\citenamefont {Artime},
  \citenamefont {Grassia}, \citenamefont {De~Domenico}, \citenamefont
  {Gleeson}, \citenamefont {Makse}, \citenamefont {Mangioni}, \citenamefont
  {Perc},\ and\ \citenamefont {Radicchi}}]{artime2024robustness}%
  \BibitemOpen
  \bibfield  {author} {\bibinfo {author} {\bibfnamefont {O.}~\bibnamefont
  {Artime}}, \bibinfo {author} {\bibfnamefont {M.}~\bibnamefont {Grassia}},
  \bibinfo {author} {\bibfnamefont {M.}~\bibnamefont {De~Domenico}}, \bibinfo
  {author} {\bibfnamefont {J.~P.}\ \bibnamefont {Gleeson}}, \bibinfo {author}
  {\bibfnamefont {H.~A.}\ \bibnamefont {Makse}}, \bibinfo {author}
  {\bibfnamefont {G.}~\bibnamefont {Mangioni}}, \bibinfo {author}
  {\bibfnamefont {M.}~\bibnamefont {Perc}},\ and\ \bibinfo {author}
  {\bibfnamefont {F.}~\bibnamefont {Radicchi}},\ }\bibfield  {title} {\bibinfo
  {title} {Robustness and resilience of complex networks},\ }\href
  {https://doi.org/https://doi.org/10.1038/s42254-023-00676-y} {\bibfield
  {journal} {\bibinfo  {journal} {Nature Reviews Physics}\ }\textbf {\bibinfo
  {volume} {6}},\ \bibinfo {pages} {114} (\bibinfo {year} {2024})}\BibitemShut
  {NoStop}%
\bibitem [{\citenamefont {Nokkala}\ \emph {et~al.}(2024)\citenamefont
  {Nokkala}, \citenamefont {Piilo},\ and\ \citenamefont
  {Bianconi}}]{nokkala2024complex}%
  \BibitemOpen
  \bibfield  {author} {\bibinfo {author} {\bibfnamefont {J.}~\bibnamefont
  {Nokkala}}, \bibinfo {author} {\bibfnamefont {J.}~\bibnamefont {Piilo}},\
  and\ \bibinfo {author} {\bibfnamefont {G.}~\bibnamefont {Bianconi}},\
  }\bibfield  {title} {\bibinfo {title} {Complex quantum networks: a topical
  review},\ }\href {https://doi.org/10.1088/1751-8121/ad41a6} {\bibfield
  {journal} {\bibinfo  {journal} {Journal of Physics A: Mathematical and
  Theoretical}\ }\textbf {\bibinfo {volume} {57}},\ \bibinfo {pages} {233001}
  (\bibinfo {year} {2024})}\BibitemShut {NoStop}%
\bibitem [{\citenamefont {Pirandola}\ \emph {et~al.}(2020)\citenamefont
  {Pirandola}, \citenamefont {Andersen}, \citenamefont {Banchi}, \citenamefont
  {Berta}, \citenamefont {Bunandar}, \citenamefont {Colbeck}, \citenamefont
  {Englund}, \citenamefont {Gehring}, \citenamefont {Lupo}, \citenamefont
  {Ottaviani}, \citenamefont {Pereira}, \citenamefont {Razavi}, \citenamefont
  {Shaari}, \citenamefont {Tomamichel}, \citenamefont {Usenko}, \citenamefont
  {Vallone}, \citenamefont {Villoresi},\ and\ \citenamefont
  {Wallden}}]{pirandola2020advances}%
  \BibitemOpen
  \bibfield  {author} {\bibinfo {author} {\bibfnamefont {S.}~\bibnamefont
  {Pirandola}}, \bibinfo {author} {\bibfnamefont {U.~L.}\ \bibnamefont
  {Andersen}}, \bibinfo {author} {\bibfnamefont {L.}~\bibnamefont {Banchi}},
  \bibinfo {author} {\bibfnamefont {M.}~\bibnamefont {Berta}}, \bibinfo
  {author} {\bibfnamefont {D.}~\bibnamefont {Bunandar}}, \bibinfo {author}
  {\bibfnamefont {R.}~\bibnamefont {Colbeck}}, \bibinfo {author} {\bibfnamefont
  {D.}~\bibnamefont {Englund}}, \bibinfo {author} {\bibfnamefont
  {T.}~\bibnamefont {Gehring}}, \bibinfo {author} {\bibfnamefont
  {C.}~\bibnamefont {Lupo}}, \bibinfo {author} {\bibfnamefont {C.}~\bibnamefont
  {Ottaviani}}, \bibinfo {author} {\bibfnamefont {J.~L.}\ \bibnamefont
  {Pereira}}, \bibinfo {author} {\bibfnamefont {M.}~\bibnamefont {Razavi}},
  \bibinfo {author} {\bibfnamefont {J.~S.}\ \bibnamefont {Shaari}}, \bibinfo
  {author} {\bibfnamefont {M.}~\bibnamefont {Tomamichel}}, \bibinfo {author}
  {\bibfnamefont {V.~C.}\ \bibnamefont {Usenko}}, \bibinfo {author}
  {\bibfnamefont {G.}~\bibnamefont {Vallone}}, \bibinfo {author} {\bibfnamefont
  {P.}~\bibnamefont {Villoresi}},\ and\ \bibinfo {author} {\bibfnamefont
  {P.}~\bibnamefont {Wallden}},\ }\bibfield  {title} {\bibinfo {title}
  {Advances in quantum cryptography},\ }\href
  {https://doi.org/10.1364/AOP.361502} {\bibfield  {journal} {\bibinfo
  {journal} {Adv. Opt. Photon.}\ }\textbf {\bibinfo {volume} {12}},\ \bibinfo
  {pages} {1012} (\bibinfo {year} {2020})}\BibitemShut {NoStop}%
\bibitem [{\citenamefont {Perseguers}\ \emph {et~al.}(2010)\citenamefont
  {Perseguers}, \citenamefont {Lewenstein}, \citenamefont {Ac\'{\i}n},\ and\
  \citenamefont {Cirac}}]{perseguers2010quantum}%
  \BibitemOpen
  \bibfield  {author} {\bibinfo {author} {\bibfnamefont {S.}~\bibnamefont
  {Perseguers}}, \bibinfo {author} {\bibfnamefont {M.}~\bibnamefont
  {Lewenstein}}, \bibinfo {author} {\bibfnamefont {A.}~\bibnamefont
  {Ac\'{\i}n}},\ and\ \bibinfo {author} {\bibfnamefont {J.~I.}\ \bibnamefont
  {Cirac}},\ }\bibfield  {title} {\bibinfo {title} {Quantum random networks},\
  }\href {https://doi.org/https://doi.org/10.1038/nphys1665} {\bibfield
  {journal} {\bibinfo  {journal} {Nature Physics}\ }\textbf {\bibinfo {volume}
  {6}},\ \bibinfo {pages} {539} (\bibinfo {year} {2010})}\BibitemShut {NoStop}%
\bibitem [{\citenamefont {Perseguers}\ \emph {et~al.}(2008)\citenamefont
  {Perseguers}, \citenamefont {Cirac}, \citenamefont {Ac\'{\i}n}, \citenamefont
  {Lewenstein},\ and\ \citenamefont {Wehr}}]{perseguers2008entanglement}%
  \BibitemOpen
  \bibfield  {author} {\bibinfo {author} {\bibfnamefont {S.}~\bibnamefont
  {Perseguers}}, \bibinfo {author} {\bibfnamefont {J.~I.}\ \bibnamefont
  {Cirac}}, \bibinfo {author} {\bibfnamefont {A.}~\bibnamefont {Ac\'{\i}n}},
  \bibinfo {author} {\bibfnamefont {M.}~\bibnamefont {Lewenstein}},\ and\
  \bibinfo {author} {\bibfnamefont {J.}~\bibnamefont {Wehr}},\ }\bibfield
  {title} {\bibinfo {title} {Entanglement distribution in pure-state quantum
  networks},\ }\href {https://doi.org/10.1103/PhysRevA.77.022308} {\bibfield
  {journal} {\bibinfo  {journal} {Phys. Rev. A}\ }\textbf {\bibinfo {volume}
  {77}},\ \bibinfo {pages} {022308} (\bibinfo {year} {2008})}\BibitemShut
  {NoStop}%
\bibitem [{\citenamefont {Coutinho}\ \emph {et~al.}(2022)\citenamefont
  {Coutinho}, \citenamefont {Munro}, \citenamefont {Nemoto},\ and\
  \citenamefont {Omar}}]{coutinho2022robustness}%
  \BibitemOpen
  \bibfield  {author} {\bibinfo {author} {\bibfnamefont {B.~C.}\ \bibnamefont
  {Coutinho}}, \bibinfo {author} {\bibfnamefont {W.~J.}\ \bibnamefont {Munro}},
  \bibinfo {author} {\bibfnamefont {K.}~\bibnamefont {Nemoto}},\ and\ \bibinfo
  {author} {\bibfnamefont {Y.}~\bibnamefont {Omar}},\ }\bibfield  {title}
  {\bibinfo {title} {Robustness of noisy quantum networks},\ }\href
  {https://doi.org/https://doi.org/10.1038/s42005-022-00866-7} {\bibfield
  {journal} {\bibinfo  {journal} {Communications Physics}\ }\textbf {\bibinfo
  {volume} {5}},\ \bibinfo {pages} {105} (\bibinfo {year} {2022})}\BibitemShut
  {NoStop}%
\bibitem [{\citenamefont {Pirandola}(2019)}]{pirandola2019end}%
  \BibitemOpen
  \bibfield  {author} {\bibinfo {author} {\bibfnamefont {S.}~\bibnamefont
  {Pirandola}},\ }\bibfield  {title} {\bibinfo {title} {End-to-end capacities
  of a quantum communication network},\ }\href
  {https://doi.org/https://doi.org/10.1038/s42005-019-0147-3} {\bibfield
  {journal} {\bibinfo  {journal} {Communications Physics}\ }\textbf {\bibinfo
  {volume} {2}},\ \bibinfo {pages} {51} (\bibinfo {year} {2019})}\BibitemShut
  {NoStop}%
\bibitem [{\citenamefont {Zwerger}\ \emph {et~al.}(2018)\citenamefont
  {Zwerger}, \citenamefont {Pirker}, \citenamefont {Dunjko}, \citenamefont
  {Briegel},\ and\ \citenamefont {D\"ur}}]{zwerger2018long}%
  \BibitemOpen
  \bibfield  {author} {\bibinfo {author} {\bibfnamefont {M.}~\bibnamefont
  {Zwerger}}, \bibinfo {author} {\bibfnamefont {A.}~\bibnamefont {Pirker}},
  \bibinfo {author} {\bibfnamefont {V.}~\bibnamefont {Dunjko}}, \bibinfo
  {author} {\bibfnamefont {H.~J.}\ \bibnamefont {Briegel}},\ and\ \bibinfo
  {author} {\bibfnamefont {W.}~\bibnamefont {D\"ur}},\ }\bibfield  {title}
  {\bibinfo {title} {Long-range big quantum-data transmission},\ }\href
  {https://doi.org/10.1103/PhysRevLett.120.030503} {\bibfield  {journal}
  {\bibinfo  {journal} {Phys. Rev. Lett.}\ }\textbf {\bibinfo {volume} {120}},\
  \bibinfo {pages} {030503} (\bibinfo {year} {2018})}\BibitemShut {NoStop}%
\bibitem [{\citenamefont {Chen}\ \emph {et~al.}(2021)\citenamefont {Chen},
  \citenamefont {Jiang}, \citenamefont {Tang}, \citenamefont {Zhou},
  \citenamefont {Yuan}, \citenamefont {Zhou}, \citenamefont {Wang},
  \citenamefont {Liu}, \citenamefont {Chen}, \citenamefont {Liu} \emph
  {et~al.}}]{chen2021implementation}%
  \BibitemOpen
  \bibfield  {author} {\bibinfo {author} {\bibfnamefont {T.-Y.}\ \bibnamefont
  {Chen}}, \bibinfo {author} {\bibfnamefont {X.}~\bibnamefont {Jiang}},
  \bibinfo {author} {\bibfnamefont {S.-B.}\ \bibnamefont {Tang}}, \bibinfo
  {author} {\bibfnamefont {L.}~\bibnamefont {Zhou}}, \bibinfo {author}
  {\bibfnamefont {X.}~\bibnamefont {Yuan}}, \bibinfo {author} {\bibfnamefont
  {H.}~\bibnamefont {Zhou}}, \bibinfo {author} {\bibfnamefont {J.}~\bibnamefont
  {Wang}}, \bibinfo {author} {\bibfnamefont {Y.}~\bibnamefont {Liu}}, \bibinfo
  {author} {\bibfnamefont {L.-K.}\ \bibnamefont {Chen}}, \bibinfo {author}
  {\bibfnamefont {W.-Y.}\ \bibnamefont {Liu}}, \emph {et~al.},\ }\bibfield
  {title} {\bibinfo {title} {Implementation of a 46-node quantum metropolitan
  area network},\ }\href
  {https://doi.org/https://doi.org/10.1038/s41534-021-00474-3} {\bibfield
  {journal} {\bibinfo  {journal} {npj Quantum Information}\ }\textbf {\bibinfo
  {volume} {7}},\ \bibinfo {pages} {134} (\bibinfo {year} {2021})}\BibitemShut
  {NoStop}%
\bibitem [{\citenamefont {Bianconi}(2018)}]{bianconi2018multilayer}%
  \BibitemOpen
  \bibfield  {author} {\bibinfo {author} {\bibfnamefont {G.}~\bibnamefont
  {Bianconi}},\ }\href {https://doi.org/10.1093/oso/9780198753919.001.0001}
  {\emph {\bibinfo {title} {{Multilayer Networks: Structure and Function}}}}\
  (\bibinfo  {publisher} {Oxford University Press},\ \bibinfo {year}
  {2018})\BibitemShut {NoStop}%
\bibitem [{\citenamefont {Cirigliano}\ \emph {et~al.}(2023)\citenamefont
  {Cirigliano}, \citenamefont {Castellano},\ and\ \citenamefont
  {Tim\'ar}}]{cirigliano2023extended}%
  \BibitemOpen
  \bibfield  {author} {\bibinfo {author} {\bibfnamefont {L.}~\bibnamefont
  {Cirigliano}}, \bibinfo {author} {\bibfnamefont {C.}~\bibnamefont
  {Castellano}},\ and\ \bibinfo {author} {\bibfnamefont {G.}~\bibnamefont
  {Tim\'ar}},\ }\bibfield  {title} {\bibinfo {title} {Extended-range
  percolation in complex networks},\ }\href
  {https://doi.org/10.1103/PhysRevE.108.044304} {\bibfield  {journal} {\bibinfo
   {journal} {Phys. Rev. E}\ }\textbf {\bibinfo {volume} {108}},\ \bibinfo
  {pages} {044304} (\bibinfo {year} {2023})}\BibitemShut {NoStop}%
\bibitem [{\citenamefont {Malarz}(2015)}]{Malarz2015}%
  \BibitemOpen
  \bibfield  {author} {\bibinfo {author} {\bibfnamefont {K.}~\bibnamefont
  {Malarz}},\ }\bibfield  {title} {\bibinfo {title} {Simple cubic random-site
  percolation thresholds for neighborhoods containing fourth-near est
  neighbors},\ }\href {https://doi.org/10.1103/PhysRevE.91.043301} {\bibfield
  {journal} {\bibinfo  {journal} {Phys. Rev. E}\ }\textbf {\bibinfo {volume}
  {91}},\ \bibinfo {pages} {043301} (\bibinfo {year} {2015})}\BibitemShut
  {NoStop}%
\bibitem [{\citenamefont {Malarz}(2020)}]{malarz2020site}%
  \BibitemOpen
  \bibfield  {author} {\bibinfo {author} {\bibfnamefont {K.}~\bibnamefont
  {Malarz}},\ }\bibfield  {title} {\bibinfo {title} {Site percolation
  thresholds on triangular lattice with complex neighborhoods},\ }\href
  {https://doi.org/10.1063/5.0022336} {\bibfield  {journal} {\bibinfo
  {journal} {Chaos}\ }\textbf {\bibinfo {volume} {30}},\ \bibinfo {pages}
  {123123} (\bibinfo {year} {2020})}\BibitemShut {NoStop}%
\bibitem [{\citenamefont {Malarz}(2021)}]{malarz2021percolation}%
  \BibitemOpen
  \bibfield  {author} {\bibinfo {author} {\bibfnamefont {K.}~\bibnamefont
  {Malarz}},\ }\bibfield  {title} {\bibinfo {title} {Percolation thresholds on
  a triangular lattice for neighborhoods containing sites up to the fifth
  coordination zone},\ }\href {https://doi.org/10.1103/PhysRevE.103.052107}
  {\bibfield  {journal} {\bibinfo  {journal} {Phys. Rev. E}\ }\textbf {\bibinfo
  {volume} {103}},\ \bibinfo {pages} {052107} (\bibinfo {year}
  {2021})}\BibitemShut {NoStop}%
\bibitem [{\citenamefont {Xun}\ and\ \citenamefont {Ziff}(2020)}]{xun2020bond}%
  \BibitemOpen
  \bibfield  {author} {\bibinfo {author} {\bibfnamefont {Z.}~\bibnamefont
  {Xun}}\ and\ \bibinfo {author} {\bibfnamefont {R.~M.}\ \bibnamefont {Ziff}},\
  }\bibfield  {title} {\bibinfo {title} {Bond percolation on simple cubic
  lattices with extended neighborhoods},\ }\href
  {https://doi.org/10.1103/PhysRevE.102.012102} {\bibfield  {journal} {\bibinfo
   {journal} {Phys. Rev. E}\ }\textbf {\bibinfo {volume} {102}},\ \bibinfo
  {pages} {012102} (\bibinfo {year} {2020})}\BibitemShut {NoStop}%
\bibitem [{\citenamefont {Xun}\ \emph {et~al.}(2021)\citenamefont {Xun},
  \citenamefont {Hao},\ and\ \citenamefont {Ziff}}]{Xun2021}%
  \BibitemOpen
  \bibfield  {author} {\bibinfo {author} {\bibfnamefont {Z.}~\bibnamefont
  {Xun}}, \bibinfo {author} {\bibfnamefont {D.}~\bibnamefont {Hao}},\ and\
  \bibinfo {author} {\bibfnamefont {R.~M.}\ \bibnamefont {Ziff}},\ }\bibfield
  {title} {\bibinfo {title} {Site percolation on square and simple cubic
  lattices with extended neighborhoods and the ir continuum limit},\ }\href
  {https://doi.org/10.1103/PhysRevE.103.022126} {\bibfield  {journal} {\bibinfo
   {journal} {Phys. Rev. E}\ }\textbf {\bibinfo {volume} {103}},\ \bibinfo
  {pages} {022126} (\bibinfo {year} {2021})}\BibitemShut {NoStop}%
\bibitem [{\citenamefont {Zhao}\ \emph {et~al.}(2022)\citenamefont {Zhao},
  \citenamefont {Yan}, \citenamefont {Xun}, \citenamefont {Hao},\ and\
  \citenamefont {Ziff}}]{zhao2022site}%
  \BibitemOpen
  \bibfield  {author} {\bibinfo {author} {\bibfnamefont {P.}~\bibnamefont
  {Zhao}}, \bibinfo {author} {\bibfnamefont {J.}~\bibnamefont {Yan}}, \bibinfo
  {author} {\bibfnamefont {Z.}~\bibnamefont {Xun}}, \bibinfo {author}
  {\bibfnamefont {D.}~\bibnamefont {Hao}},\ and\ \bibinfo {author}
  {\bibfnamefont {R.~M.}\ \bibnamefont {Ziff}},\ }\bibfield  {title} {\bibinfo
  {title} {Site and bond percolation on four-dimensional simple hypercubic
  lattices with extended neighborhoods},\ }\href
  {https://doi.org/10.1088/1742-5468/ac52a8} {\bibfield  {journal} {\bibinfo
  {journal} {J. Stat. Mech. Theory Exp.}\ }\textbf {\bibinfo {volume} {2022}},\
  \bibinfo {pages} {033202} (\bibinfo {year} {2022})}\BibitemShut {NoStop}%
\bibitem [{\citenamefont {Xun}\ \emph {et~al.}(2022)\citenamefont {Xun},
  \citenamefont {Hao},\ and\ \citenamefont {Ziff}}]{xun2022site}%
  \BibitemOpen
  \bibfield  {author} {\bibinfo {author} {\bibfnamefont {Z.}~\bibnamefont
  {Xun}}, \bibinfo {author} {\bibfnamefont {D.}~\bibnamefont {Hao}},\ and\
  \bibinfo {author} {\bibfnamefont {R.~M.}\ \bibnamefont {Ziff}},\ }\bibfield
  {title} {\bibinfo {title} {Site and bond percolation thresholds on regular
  lattices with compact extended-range neighborhoods in two and three
  dimensions},\ }\href {https://doi.org/10.1103/PhysRevE.105.024105} {\bibfield
   {journal} {\bibinfo  {journal} {Phys. Rev. E}\ }\textbf {\bibinfo {volume}
  {105}},\ \bibinfo {pages} {024105} (\bibinfo {year} {2022})}\BibitemShut
  {NoStop}%
\bibitem [{\citenamefont {Meng}\ \emph {et~al.}(2021)\citenamefont {Meng},
  \citenamefont {Gao},\ and\ \citenamefont {Havlin}}]{meng2021concurrence}%
  \BibitemOpen
  \bibfield  {author} {\bibinfo {author} {\bibfnamefont {X.}~\bibnamefont
  {Meng}}, \bibinfo {author} {\bibfnamefont {J.}~\bibnamefont {Gao}},\ and\
  \bibinfo {author} {\bibfnamefont {S.}~\bibnamefont {Havlin}},\ }\bibfield
  {title} {\bibinfo {title} {Concurrence percolation in quantum networks},\
  }\href {https://doi.org/10.1103/PhysRevLett.126.170501} {\bibfield  {journal}
  {\bibinfo  {journal} {Phys. Rev. Lett.}\ }\textbf {\bibinfo {volume} {126}},\
  \bibinfo {pages} {170501} (\bibinfo {year} {2021})}\BibitemShut {NoStop}%
\bibitem [{\citenamefont {Kim}\ and\ \citenamefont
  {Radicchi}(2024)}]{kim2024shortest}%
  \BibitemOpen
  \bibfield  {author} {\bibinfo {author} {\bibfnamefont {M.}~\bibnamefont
  {Kim}}\ and\ \bibinfo {author} {\bibfnamefont {F.}~\bibnamefont {Radicchi}},\
  }\bibfield  {title} {\bibinfo {title} {Shortest-path percolation on random
  networks},\ }\href {https://doi.org/10.1103/PhysRevLett.133.047402}
  {\bibfield  {journal} {\bibinfo  {journal} {Phys. Rev. Lett.}\ }\textbf
  {\bibinfo {volume} {133}},\ \bibinfo {pages} {047402} (\bibinfo {year}
  {2024})}\BibitemShut {NoStop}%
\bibitem [{\citenamefont {Min}\ \emph {et~al.}(2024)\citenamefont {Min},
  \citenamefont {Park}, \citenamefont {Gwak},\ and\ \citenamefont
  {Goh}}]{min2024no}%
  \BibitemOpen
  \bibfield  {author} {\bibinfo {author} {\bibfnamefont {B.}~\bibnamefont
  {Min}}, \bibinfo {author} {\bibfnamefont {E.-K.}\ \bibnamefont {Park}},
  \bibinfo {author} {\bibfnamefont {S.-H.}\ \bibnamefont {Gwak}},\ and\
  \bibinfo {author} {\bibfnamefont {K.-I.}\ \bibnamefont {Goh}},\ }\bibfield
  {title} {\bibinfo {title} {No-exclaves percolation on random networks},\
  }\href {https://doi.org/https://doi.org/10.1016/j.chaos.2024.115004}
  {\bibfield  {journal} {\bibinfo  {journal} {Chaos, Solitons \& Fractals}\
  }\textbf {\bibinfo {volume} {184}},\ \bibinfo {pages} {115004} (\bibinfo
  {year} {2024})}\BibitemShut {NoStop}%
\bibitem [{\citenamefont {Krause}\ \emph {et~al.}(2016)\citenamefont {Krause},
  \citenamefont {Danziger},\ and\ \citenamefont {Zlati\ifmmode~\acute{c}\else
  \'{c}\fi{}}}]{krause2016hidden}%
  \BibitemOpen
  \bibfield  {author} {\bibinfo {author} {\bibfnamefont {S.~M.}\ \bibnamefont
  {Krause}}, \bibinfo {author} {\bibfnamefont {M.~M.}\ \bibnamefont
  {Danziger}},\ and\ \bibinfo {author} {\bibfnamefont {V.}~\bibnamefont
  {Zlati\ifmmode~\acute{c}\else \'{c}\fi{}}},\ }\bibfield  {title} {\bibinfo
  {title} {Hidden connectivity in networks with vulnerable classes of nodes},\
  }\href {https://doi.org/10.1103/PhysRevX.6.041022} {\bibfield  {journal}
  {\bibinfo  {journal} {Phys. Rev. X}\ }\textbf {\bibinfo {volume} {6}},\
  \bibinfo {pages} {041022} (\bibinfo {year} {2016})}\BibitemShut {NoStop}%
\bibitem [{\citenamefont {Krause}\ \emph {et~al.}(2017)\citenamefont {Krause},
  \citenamefont {Danziger},\ and\ \citenamefont {Zlati\ifmmode~\acute{c}\else
  \'{c}\fi{}}}]{krause2017color}%
  \BibitemOpen
  \bibfield  {author} {\bibinfo {author} {\bibfnamefont {S.~M.}\ \bibnamefont
  {Krause}}, \bibinfo {author} {\bibfnamefont {M.~M.}\ \bibnamefont
  {Danziger}},\ and\ \bibinfo {author} {\bibfnamefont {V.}~\bibnamefont
  {Zlati\ifmmode~\acute{c}\else \'{c}\fi{}}},\ }\bibfield  {title} {\bibinfo
  {title} {Color-avoiding percolation},\ }\href
  {https://doi.org/10.1103/PhysRevE.96.022313} {\bibfield  {journal} {\bibinfo
  {journal} {Phys. Rev. E}\ }\textbf {\bibinfo {volume} {96}},\ \bibinfo
  {pages} {022313} (\bibinfo {year} {2017})}\BibitemShut {NoStop}%
\bibitem [{\citenamefont {Castellano}\ and\ \citenamefont
  {Pastor-Satorras}(2020)}]{castellano2020cumulative}%
  \BibitemOpen
  \bibfield  {author} {\bibinfo {author} {\bibfnamefont {C.}~\bibnamefont
  {Castellano}}\ and\ \bibinfo {author} {\bibfnamefont {R.}~\bibnamefont
  {Pastor-Satorras}},\ }\bibfield  {title} {\bibinfo {title} {Cumulative
  merging percolation and the epidemic transition of the
  susceptible-infected-susceptible model in networks},\ }\href
  {https://doi.org/10.1103/PhysRevX.10.011070} {\bibfield  {journal} {\bibinfo
  {journal} {Phys. Rev. X}\ }\textbf {\bibinfo {volume} {10}},\ \bibinfo
  {pages} {011070} (\bibinfo {year} {2020})}\BibitemShut {NoStop}%
\bibitem [{\citenamefont {Cirigliano}\ \emph {et~al.}(2022)\citenamefont
  {Cirigliano}, \citenamefont {Cimini}, \citenamefont {Pastor-Satorras},\ and\
  \citenamefont {Castellano}}]{cirigliano2022cumulative}%
  \BibitemOpen
  \bibfield  {author} {\bibinfo {author} {\bibfnamefont {L.}~\bibnamefont
  {Cirigliano}}, \bibinfo {author} {\bibfnamefont {G.}~\bibnamefont {Cimini}},
  \bibinfo {author} {\bibfnamefont {R.}~\bibnamefont {Pastor-Satorras}},\ and\
  \bibinfo {author} {\bibfnamefont {C.}~\bibnamefont {Castellano}},\ }\bibfield
   {title} {\bibinfo {title} {Cumulative merging percolation: A long-range
  percolation process in networks},\ }\href
  {https://doi.org/10.1103/PhysRevE.105.054310} {\bibfield  {journal} {\bibinfo
   {journal} {Phys. Rev. E}\ }\textbf {\bibinfo {volume} {105}},\ \bibinfo
  {pages} {054310} (\bibinfo {year} {2022})}\BibitemShut {NoStop}%
\bibitem [{\citenamefont {Newman}(2023)}]{newman2023message}%
  \BibitemOpen
  \bibfield  {author} {\bibinfo {author} {\bibfnamefont {M.}~\bibnamefont
  {Newman}},\ }\bibfield  {title} {\bibinfo {title} {Message passing methods on
  complex networks},\ }\href
  {https://doi.org/https://doi.org/10.1098/rspa.2022.0774} {\bibfield
  {journal} {\bibinfo  {journal} {Proceedings of the Royal Society A}\ }\textbf
  {\bibinfo {volume} {479}},\ \bibinfo {pages} {20220774} (\bibinfo {year}
  {2023})}\BibitemShut {NoStop}%
\bibitem [{\citenamefont {Mézard}\ and\ \citenamefont
  {Montanari}(2009)}]{mezard2009information}%
  \BibitemOpen
  \bibfield  {author} {\bibinfo {author} {\bibfnamefont {M.}~\bibnamefont
  {Mézard}}\ and\ \bibinfo {author} {\bibfnamefont {A.}~\bibnamefont
  {Montanari}},\ }\href
  {https://doi.org/10.1093/acprof:oso/9780198570837.001.0001} {\emph {\bibinfo
  {title} {{Information, Physics, and Computation}}}}\ (\bibinfo  {publisher}
  {Oxford University Press},\ \bibinfo {year} {2009})\BibitemShut {NoStop}%
\bibitem [{\citenamefont {Liu}\ \emph {et~al.}(2011)\citenamefont {Liu},
  \citenamefont {Slotine},\ and\ \citenamefont
  {Barab{\'a}si}}]{liu2011controllability}%
  \BibitemOpen
  \bibfield  {author} {\bibinfo {author} {\bibfnamefont {Y.-Y.}\ \bibnamefont
  {Liu}}, \bibinfo {author} {\bibfnamefont {J.-J.}\ \bibnamefont {Slotine}},\
  and\ \bibinfo {author} {\bibfnamefont {A.-L.}\ \bibnamefont {Barab{\'a}si}},\
  }\bibfield  {title} {\bibinfo {title} {Controllability of complex networks},\
  }\href {https://doi.org/https://doi.org/10.1038/nature10011} {\bibfield
  {journal} {\bibinfo  {journal} {nature}\ }\textbf {\bibinfo {volume} {473}},\
  \bibinfo {pages} {167} (\bibinfo {year} {2011})}\BibitemShut {NoStop}%
\bibitem [{\citenamefont {Karrer}\ \emph {et~al.}(2014)\citenamefont {Karrer},
  \citenamefont {Newman},\ and\ \citenamefont
  {Zdeborov\'a}}]{karrer2014percolation}%
  \BibitemOpen
  \bibfield  {author} {\bibinfo {author} {\bibfnamefont {B.}~\bibnamefont
  {Karrer}}, \bibinfo {author} {\bibfnamefont {M.~E.~J.}\ \bibnamefont
  {Newman}},\ and\ \bibinfo {author} {\bibfnamefont {L.}~\bibnamefont
  {Zdeborov\'a}},\ }\bibfield  {title} {\bibinfo {title} {Percolation on sparse
  networks},\ }\href {https://doi.org/10.1103/PhysRevLett.113.208702}
  {\bibfield  {journal} {\bibinfo  {journal} {Phys. Rev. Lett.}\ }\textbf
  {\bibinfo {volume} {113}},\ \bibinfo {pages} {208702} (\bibinfo {year}
  {2014})}\BibitemShut {NoStop}%
\bibitem [{\citenamefont {Menichetti}\ \emph {et~al.}(2014)\citenamefont
  {Menichetti}, \citenamefont {Dall'Asta},\ and\ \citenamefont
  {Bianconi}}]{menichetti2014network}%
  \BibitemOpen
  \bibfield  {author} {\bibinfo {author} {\bibfnamefont {G.}~\bibnamefont
  {Menichetti}}, \bibinfo {author} {\bibfnamefont {L.}~\bibnamefont
  {Dall'Asta}},\ and\ \bibinfo {author} {\bibfnamefont {G.}~\bibnamefont
  {Bianconi}},\ }\bibfield  {title} {\bibinfo {title} {Network controllability
  is determined by the density of low in-degree and out-degree nodes},\ }\href
  {https://doi.org/10.1103/PhysRevLett.113.078701} {\bibfield  {journal}
  {\bibinfo  {journal} {Phys. Rev. Lett.}\ }\textbf {\bibinfo {volume} {113}},\
  \bibinfo {pages} {078701} (\bibinfo {year} {2014})}\BibitemShut {NoStop}%
\bibitem [{\citenamefont {Radicchi}\ and\ \citenamefont
  {Bianconi}(2017)}]{radicchi2017redundant}%
  \BibitemOpen
  \bibfield  {author} {\bibinfo {author} {\bibfnamefont {F.}~\bibnamefont
  {Radicchi}}\ and\ \bibinfo {author} {\bibfnamefont {G.}~\bibnamefont
  {Bianconi}},\ }\bibfield  {title} {\bibinfo {title} {Redundant
  interdependencies boost the robustness of multiplex networks},\ }\href
  {https://doi.org/10.1103/PhysRevX.7.011013} {\bibfield  {journal} {\bibinfo
  {journal} {Phys. Rev. X}\ }\textbf {\bibinfo {volume} {7}},\ \bibinfo {pages}
  {011013} (\bibinfo {year} {2017})}\BibitemShut {NoStop}%
\bibitem [{\citenamefont {Zdeborov\'a}\ and\ \citenamefont
  {Krz\k{a}ka\l{}a}(2007)}]{zdeborova2007phase}%
  \BibitemOpen
  \bibfield  {author} {\bibinfo {author} {\bibfnamefont {L.}~\bibnamefont
  {Zdeborov\'a}}\ and\ \bibinfo {author} {\bibfnamefont {F.}~\bibnamefont
  {Krz\k{a}ka\l{}a}},\ }\bibfield  {title} {\bibinfo {title} {Phase transitions
  in the coloring of random graphs},\ }\href
  {https://doi.org/10.1103/PhysRevE.76.031131} {\bibfield  {journal} {\bibinfo
  {journal} {Phys. Rev. E}\ }\textbf {\bibinfo {volume} {76}},\ \bibinfo
  {pages} {031131} (\bibinfo {year} {2007})}\BibitemShut {NoStop}%
\bibitem [{\citenamefont {Hartmann}\ and\ \citenamefont
  {Weigt}(2006)}]{hartmann2006phase}%
  \BibitemOpen
  \bibfield  {author} {\bibinfo {author} {\bibfnamefont {A.~K.}\ \bibnamefont
  {Hartmann}}\ and\ \bibinfo {author} {\bibfnamefont {M.}~\bibnamefont
  {Weigt}},\ }\href
  {https://www.wiley.com/en-us/Phase+Transitions+in+Combinatorial+Optimization+Problems%3A+Basics%2C+Algorithms+and+Statistical+Mechanics-p-9783527606733}
  {\emph {\bibinfo {title} {Phase transitions in combinatorial optimization
  problems: basics, algorithms and statistical mechanics}}}\ (\bibinfo
  {publisher} {John Wiley \& Sons},\ \bibinfo {year} {2006})\BibitemShut
  {NoStop}%
\bibitem [{\citenamefont {Cantwell}\ \emph {et~al.}(2023)\citenamefont
  {Cantwell}, \citenamefont {Kirkley},\ and\ \citenamefont
  {Radicchi}}]{cantwell2023heterogeneous}%
  \BibitemOpen
  \bibfield  {author} {\bibinfo {author} {\bibfnamefont {G.~T.}\ \bibnamefont
  {Cantwell}}, \bibinfo {author} {\bibfnamefont {A.}~\bibnamefont {Kirkley}},\
  and\ \bibinfo {author} {\bibfnamefont {F.}~\bibnamefont {Radicchi}},\
  }\bibfield  {title} {\bibinfo {title} {Heterogeneous message passing for
  heterogeneous networks},\ }\href
  {https://doi.org/10.1103/PhysRevE.108.034310} {\bibfield  {journal} {\bibinfo
   {journal} {Phys. Rev. E}\ }\textbf {\bibinfo {volume} {108}},\ \bibinfo
  {pages} {034310} (\bibinfo {year} {2023})}\BibitemShut {NoStop}%
\bibitem [{\citenamefont {Kirkley}\ \emph {et~al.}(2021)\citenamefont
  {Kirkley}, \citenamefont {Cantwell},\ and\ \citenamefont
  {Newman}}]{kirkley2021belief}%
  \BibitemOpen
  \bibfield  {author} {\bibinfo {author} {\bibfnamefont {A.}~\bibnamefont
  {Kirkley}}, \bibinfo {author} {\bibfnamefont {G.~T.}\ \bibnamefont
  {Cantwell}},\ and\ \bibinfo {author} {\bibfnamefont {M.~E.~J.}\ \bibnamefont
  {Newman}},\ }\bibfield  {title} {\bibinfo {title} {Belief propagation for
  networks with loops},\ }\href {https://doi.org/10.1126/sciadv.abf1211}
  {\bibfield  {journal} {\bibinfo  {journal} {Science Advances}\ }\textbf
  {\bibinfo {volume} {7}},\ \bibinfo {pages} {eabf1211} (\bibinfo {year}
  {2021})}\BibitemShut {NoStop}%
\bibitem [{\citenamefont {Boccaletti}\ \emph {et~al.}(2014)\citenamefont
  {Boccaletti}, \citenamefont {Bianconi}, \citenamefont {Criado}, \citenamefont
  {{del Genio}}, \citenamefont {Gómez-Gardeñes}, \citenamefont {Romance},
  \citenamefont {Sendiña-Nadal}, \citenamefont {Wang},\ and\ \citenamefont
  {Zanin}}]{boccaletti2014structure}%
  \BibitemOpen
  \bibfield  {author} {\bibinfo {author} {\bibfnamefont {S.}~\bibnamefont
  {Boccaletti}}, \bibinfo {author} {\bibfnamefont {G.}~\bibnamefont
  {Bianconi}}, \bibinfo {author} {\bibfnamefont {R.}~\bibnamefont {Criado}},
  \bibinfo {author} {\bibfnamefont {C.}~\bibnamefont {{del Genio}}}, \bibinfo
  {author} {\bibfnamefont {J.}~\bibnamefont {Gómez-Gardeñes}}, \bibinfo
  {author} {\bibfnamefont {M.}~\bibnamefont {Romance}}, \bibinfo {author}
  {\bibfnamefont {I.}~\bibnamefont {Sendiña-Nadal}}, \bibinfo {author}
  {\bibfnamefont {Z.}~\bibnamefont {Wang}},\ and\ \bibinfo {author}
  {\bibfnamefont {M.}~\bibnamefont {Zanin}},\ }\bibfield  {title} {\bibinfo
  {title} {The structure and dynamics of multilayer networks},\ }\href
  {https://doi.org/https://doi.org/10.1016/j.physrep.2014.07.001} {\bibfield
  {journal} {\bibinfo  {journal} {Physics Reports}\ }\textbf {\bibinfo {volume}
  {544}},\ \bibinfo {pages} {1} (\bibinfo {year} {2014})}\BibitemShut {NoStop}%
\bibitem [{\citenamefont {Kivelä}\ \emph {et~al.}(2014)\citenamefont
  {Kivelä}, \citenamefont {Arenas}, \citenamefont {Barthelemy}, \citenamefont
  {Gleeson}, \citenamefont {Moreno},\ and\ \citenamefont
  {Porter}}]{kivela2014multilayer}%
  \BibitemOpen
  \bibfield  {author} {\bibinfo {author} {\bibfnamefont {M.}~\bibnamefont
  {Kivelä}}, \bibinfo {author} {\bibfnamefont {A.}~\bibnamefont {Arenas}},
  \bibinfo {author} {\bibfnamefont {M.}~\bibnamefont {Barthelemy}}, \bibinfo
  {author} {\bibfnamefont {J.~P.}\ \bibnamefont {Gleeson}}, \bibinfo {author}
  {\bibfnamefont {Y.}~\bibnamefont {Moreno}},\ and\ \bibinfo {author}
  {\bibfnamefont {M.~A.}\ \bibnamefont {Porter}},\ }\bibfield  {title}
  {\bibinfo {title} {{Multilayer networks}},\ }\href
  {https://doi.org/10.1093/comnet/cnu016} {\bibfield  {journal} {\bibinfo
  {journal} {Journal of Complex Networks}\ }\textbf {\bibinfo {volume} {2}},\
  \bibinfo {pages} {203} (\bibinfo {year} {2014})}\BibitemShut {NoStop}%
\bibitem [{\citenamefont {Buldyrev}\ \emph {et~al.}(2010)\citenamefont
  {Buldyrev}, \citenamefont {Parshani}, \citenamefont {Paul}, \citenamefont
  {Stanley},\ and\ \citenamefont {Havlin}}]{buldyrev2010catastrophic}%
  \BibitemOpen
  \bibfield  {author} {\bibinfo {author} {\bibfnamefont {S.~V.}\ \bibnamefont
  {Buldyrev}}, \bibinfo {author} {\bibfnamefont {R.}~\bibnamefont {Parshani}},
  \bibinfo {author} {\bibfnamefont {G.}~\bibnamefont {Paul}}, \bibinfo {author}
  {\bibfnamefont {H.~E.}\ \bibnamefont {Stanley}},\ and\ \bibinfo {author}
  {\bibfnamefont {S.}~\bibnamefont {Havlin}},\ }\bibfield  {title} {\bibinfo
  {title} {Catastrophic cascade of failures in interdependent networks},\
  }\href {https://doi.org/https://doi.org/10.1038/nature08932} {\bibfield
  {journal} {\bibinfo  {journal} {Nature}\ }\textbf {\bibinfo {volume} {464}},\
  \bibinfo {pages} {1025} (\bibinfo {year} {2010})}\BibitemShut {NoStop}%
\bibitem [{\citenamefont {Baxter}\ \emph {et~al.}(2012)\citenamefont {Baxter},
  \citenamefont {Dorogovtsev}, \citenamefont {Goltsev},\ and\ \citenamefont
  {Mendes}}]{baxter2012avalanche}%
  \BibitemOpen
  \bibfield  {author} {\bibinfo {author} {\bibfnamefont {G.~J.}\ \bibnamefont
  {Baxter}}, \bibinfo {author} {\bibfnamefont {S.~N.}\ \bibnamefont
  {Dorogovtsev}}, \bibinfo {author} {\bibfnamefont {A.~V.}\ \bibnamefont
  {Goltsev}},\ and\ \bibinfo {author} {\bibfnamefont {J.~F.~F.}\ \bibnamefont
  {Mendes}},\ }\bibfield  {title} {\bibinfo {title} {Avalanche collapse of
  interdependent networks},\ }\href
  {https://doi.org/10.1103/PhysRevLett.109.248701} {\bibfield  {journal}
  {\bibinfo  {journal} {Phys. Rev. Lett.}\ }\textbf {\bibinfo {volume} {109}},\
  \bibinfo {pages} {248701} (\bibinfo {year} {2012})}\BibitemShut {NoStop}%
\bibitem [{\citenamefont {Cellai}\ \emph {et~al.}(2016)\citenamefont {Cellai},
  \citenamefont {Dorogovtsev},\ and\ \citenamefont
  {Bianconi}}]{cellai2016message}%
  \BibitemOpen
  \bibfield  {author} {\bibinfo {author} {\bibfnamefont {D.}~\bibnamefont
  {Cellai}}, \bibinfo {author} {\bibfnamefont {S.~N.}\ \bibnamefont
  {Dorogovtsev}},\ and\ \bibinfo {author} {\bibfnamefont {G.}~\bibnamefont
  {Bianconi}},\ }\bibfield  {title} {\bibinfo {title} {Message passing theory
  for percolation models on multiplex networks with link overlap},\ }\href
  {https://doi.org/10.1103/PhysRevE.94.032301} {\bibfield  {journal} {\bibinfo
  {journal} {Phys. Rev. E}\ }\textbf {\bibinfo {volume} {94}},\ \bibinfo
  {pages} {032301} (\bibinfo {year} {2016})}\BibitemShut {NoStop}%
\bibitem [{\citenamefont {Min}\ \emph {et~al.}(2015)\citenamefont {Min},
  \citenamefont {Lee}, \citenamefont {Lee},\ and\ \citenamefont
  {Goh}}]{min2015link}%
  \BibitemOpen
  \bibfield  {author} {\bibinfo {author} {\bibfnamefont {B.}~\bibnamefont
  {Min}}, \bibinfo {author} {\bibfnamefont {S.}~\bibnamefont {Lee}}, \bibinfo
  {author} {\bibfnamefont {K.-M.}\ \bibnamefont {Lee}},\ and\ \bibinfo {author}
  {\bibfnamefont {K.-I.}\ \bibnamefont {Goh}},\ }\bibfield  {title} {\bibinfo
  {title} {Link overlap, viability, and mutual percolation in multiplex
  networks},\ }\href
  {https://doi.org/https://doi.org/10.1016/j.chaos.2014.12.016} {\bibfield
  {journal} {\bibinfo  {journal} {Chaos, Solitons \& Fractals}\ }\textbf
  {\bibinfo {volume} {72}},\ \bibinfo {pages} {49} (\bibinfo {year}
  {2015})}\BibitemShut {NoStop}%
\bibitem [{\citenamefont {Bianconi}(2013)}]{bianconi2013statistical}%
  \BibitemOpen
  \bibfield  {author} {\bibinfo {author} {\bibfnamefont {G.}~\bibnamefont
  {Bianconi}},\ }\bibfield  {title} {\bibinfo {title} {Statistical mechanics of
  multiplex networks: Entropy and overlap},\ }\href
  {https://doi.org/10.1103/PhysRevE.87.062806} {\bibfield  {journal} {\bibinfo
  {journal} {Phys. Rev. E}\ }\textbf {\bibinfo {volume} {87}},\ \bibinfo
  {pages} {062806} (\bibinfo {year} {2013})}\BibitemShut {NoStop}%
\bibitem [{\citenamefont {Son}\ \emph {et~al.}(2012)\citenamefont {Son},
  \citenamefont {Bizhani}, \citenamefont {Christensen}, \citenamefont
  {Grassberger},\ and\ \citenamefont {Paczuski}}]{son2012percolation}%
  \BibitemOpen
  \bibfield  {author} {\bibinfo {author} {\bibfnamefont {S.-W.}\ \bibnamefont
  {Son}}, \bibinfo {author} {\bibfnamefont {G.}~\bibnamefont {Bizhani}},
  \bibinfo {author} {\bibfnamefont {C.}~\bibnamefont {Christensen}}, \bibinfo
  {author} {\bibfnamefont {P.}~\bibnamefont {Grassberger}},\ and\ \bibinfo
  {author} {\bibfnamefont {M.}~\bibnamefont {Paczuski}},\ }\bibfield  {title}
  {\bibinfo {title} {Percolation theory on interdependent networks based on
  epidemic spreading},\ }\href {https://doi.org/10.1209/0295-5075/97/16006}
  {\bibfield  {journal} {\bibinfo  {journal} {Europhysics Letters}\ }\textbf
  {\bibinfo {volume} {97}},\ \bibinfo {pages} {16006} (\bibinfo {year}
  {2012})}\BibitemShut {NoStop}%
\bibitem [{\citenamefont {Bianconi}\ and\ \citenamefont
  {Dorogovtsev}(2024)}]{bianconi2024theory}%
  \BibitemOpen
  \bibfield  {author} {\bibinfo {author} {\bibfnamefont {G.}~\bibnamefont
  {Bianconi}}\ and\ \bibinfo {author} {\bibfnamefont {S.~N.}\ \bibnamefont
  {Dorogovtsev}},\ }\bibfield  {title} {\bibinfo {title} {Theory of percolation
  on hypergraphs},\ }\href {https://doi.org/10.1103/PhysRevE.109.014306}
  {\bibfield  {journal} {\bibinfo  {journal} {Phys. Rev. E}\ }\textbf {\bibinfo
  {volume} {109}},\ \bibinfo {pages} {014306} (\bibinfo {year}
  {2024})}\BibitemShut {NoStop}%
\bibitem [{\citenamefont {Kim}\ and\ \citenamefont
  {Goh}(2024)}]{kim2024higher}%
  \BibitemOpen
  \bibfield  {author} {\bibinfo {author} {\bibfnamefont {J.-H.}\ \bibnamefont
  {Kim}}\ and\ \bibinfo {author} {\bibfnamefont {K.-I.}\ \bibnamefont {Goh}},\
  }\bibfield  {title} {\bibinfo {title} {Higher-order components dictate
  higher-order contagion dynamics in hypergraphs},\ }\href
  {https://doi.org/10.1103/PhysRevLett.132.087401} {\bibfield  {journal}
  {\bibinfo  {journal} {Phys. Rev. Lett.}\ }\textbf {\bibinfo {volume} {132}},\
  \bibinfo {pages} {087401} (\bibinfo {year} {2024})}\BibitemShut {NoStop}%
\bibitem [{\citenamefont {Sun}\ and\ \citenamefont
  {Bianconi}(2021)}]{sun2021higher}%
  \BibitemOpen
  \bibfield  {author} {\bibinfo {author} {\bibfnamefont {H.}~\bibnamefont
  {Sun}}\ and\ \bibinfo {author} {\bibfnamefont {G.}~\bibnamefont {Bianconi}},\
  }\bibfield  {title} {\bibinfo {title} {Higher-order percolation processes on
  multiplex hypergraphs},\ }\href {https://doi.org/10.1103/PhysRevE.104.034306}
  {\bibfield  {journal} {\bibinfo  {journal} {Phys. Rev. E}\ }\textbf {\bibinfo
  {volume} {104}},\ \bibinfo {pages} {034306} (\bibinfo {year}
  {2021})}\BibitemShut {NoStop}%
\bibitem [{\citenamefont {Sun}\ \emph {et~al.}(2023)\citenamefont {Sun},
  \citenamefont {Radicchi}, \citenamefont {Kurths},\ and\ \citenamefont
  {Bianconi}}]{sun2023dynamic}%
  \BibitemOpen
  \bibfield  {author} {\bibinfo {author} {\bibfnamefont {H.}~\bibnamefont
  {Sun}}, \bibinfo {author} {\bibfnamefont {F.}~\bibnamefont {Radicchi}},
  \bibinfo {author} {\bibfnamefont {J.}~\bibnamefont {Kurths}},\ and\ \bibinfo
  {author} {\bibfnamefont {G.}~\bibnamefont {Bianconi}},\ }\bibfield  {title}
  {\bibinfo {title} {The dynamic nature of percolation on networks with triadic
  interactions},\ }\href
  {https://doi.org/https://doi.org/10.1038/s41467-023-37019-5} {\bibfield
  {journal} {\bibinfo  {journal} {Nature Communications}\ }\textbf {\bibinfo
  {volume} {14}},\ \bibinfo {pages} {1308} (\bibinfo {year}
  {2023})}\BibitemShut {NoStop}%
\end{thebibliography}%

\end{document}